\documentclass[10pt,conference, compsocconf]{IEEEtran}

\usepackage[svgnames, x11names]{xcolor}

\usepackage[normalem]{ulem} % required for strikeout font

\usepackage{listings}
\lstdefinelanguage{XML}
{
basicstyle=\ttfamily\footnotesize,
  morestring=[b]",
  moredelim=[s][\bfseries\color{Maroon}]{<}{\ },
  moredelim=[s][\bfseries\color{Maroon}]{</}{>},
  moredelim=[l][\bfseries\color{Maroon}]{/>},
  moredelim=[l][\bfseries\color{Maroon}]{>},
  morecomment=[s]{<?}{?>},
  morecomment=[s]{<!--}{-->},
  commentstyle=\color{gray},
  stringstyle=\color{blue},
  identifierstyle=\color{red}
}
\usepackage[pdftex]{graphicx}
\graphicspath{{./figures/}}
\DeclareGraphicsExtensions{.pdf}

\usepackage[cmex10]{amsmath}
\usepackage{amssymb}
\usepackage{mathtools}

\usepackage{subfig}

\usepackage{algorithmicx}
\usepackage{algpseudocode}
\usepackage[ruled]{algorithm}
\definecolor{light-gray}{gray}{0.75}
\algrenewcommand{\algorithmiccomment}[1]{\hskip3em{{\footnotesize \textcolor{light-gray}{$\blacktriangleright$}}} #1}

\usepackage{multirow}

\usepackage{hyperref}

\usepackage{xspace}
\usepackage[nocompress]{cite}

\usepackage{enumitem}

\usepackage{blindtext}

\begin{document}
\title{A Meta-graph Approach to Analyze Subgraph-centric Distributed Programming Models}

\author{\IEEEauthorblockN{Ravikant Dindokar, Neel Choudhury, Yogesh Simmhan}
\IEEEauthorblockA{Department of Computational and Data Sciences,\\ Indian Institute of Science, Bangalore, India\\
ravikant7@grads.cds.iisc.ac.in, simmhan@cds.iisc.ac.in}
}

\maketitle

\begin{abstract}
Component-centric distributed graph processing platforms that use a bulk synchronous parallel (BSP) programming model have gained traction. These address the short-comings of Big Data abstractions/platforms like MapReduce/Hadoop for large-scale graph processing. However, there is limited literature on foundational aspects of the behavior of these component-centric abstractions for different graphs, graph partitioning, and graph algorithms. Here, we propose a analytical approach based on a \emph{meta-graph sketch} to examine the characteristics of component-centric graph programming models at a coarse granularity. In particular, we apply this sketch to subgraph- and block-centric abstractions, and draw a comparison with vertex-centric models like Google's Pregel. First, we explore the impact of various graph partitioning techniques on the meta-graph, and next consider the impact of the meta-graph on graph algorithms. This decouples the unwieldy large graph and their partitioning specific artifacts from their algorithmic analysis. We use 5 spatial and powerlaw graphs as exemplars, four different partitioning strategies, and PageRank and Breadth First Search as canonical algorithms. These analysis over the meta-graphs provide a reliable measure of the expected number of supersteps, and the communication and computational complexity of the algorithms for various graphs, and the relative merits of subgraph-centric models over vertex-centric ones.
\end{abstract}
%
% \section{TODO}
% \ysnote{Verify Giraph USRN. Run EURN on Giraph PR and BFS. Run EURN on GoFFish PR and BFS. Sanity check run of SSSP on other 2 source vertices for Giraph.}

% \ysnote{Analysis of results. Explain outliers.}

% \ysnote{More meta-graph statistics to explain results?}

\section{Introduction}%~~~\note{1.25 pages}}
\label{sec:intro}
%\Note{Related work and result analysis sections do not evidence the paper's contributions [Link contribution to results, related work]}\\

Distributed graph processing platforms have gained attention off-late as emerging domains such as social networks, smart transportation and deep learning grapple with large-scale graph datasets. Such domains have limited access to high-end HPC hardware, and instead leverage commodity clusters and Clouds for Big Data processing. %As such, parallel programming constructs like MPI and OpenMP give way to distributed programming models like MapReduce on such commodity hardware.
MapReduce is also less suited for graph computation, given the iterative nature of these algorithms that causes repetitive I/O~\cite{graph-twiddling}. % typically require iterative operations to perform traversals~\cite{haloop} and this forces multiple MapReduce jobs to perform repeated reads and writes of the graph structure and state to HDFS. Further, the tuple-centric view of MapReduce does easily map to a graph structure and associated programming primitives.
%
%There has also been work on distributed graph engines that offer a shared-memory perspective for irregular applications~\cite{trinity}, and on asynchronous graph-parallel computation in the form of GraphLab~\cite{graphlab}. 

Component-centric distributed graph processing platforms~\cite{goffish,blogel,giraph++} inspired by \emph{Google's Pregel}~\cite{pregel} have gained attraction due to their simplicity and scalability. Pregel offers a vertex-parallel computation model where application logic is written from the perspective of a single vertex. The logic can perform computation on the vertex and pass messages to neighboring vertices. The execution proceeds in a Bulk Synchronous Parallel (BSP) manner, where vertex computation and bulk messaging -- the pair of which forms a \emph{superstep} -- alternate and iterate until the algorithm terminates. 
%bspvaliant

While the vertex-computation itself executes in an embarrassingly data parallel manner on all vertices, the fine-grained Pregel model has communication complexity that is often $\mathcal{O}($edges$)$, and whose coordination overhead, as measured by the number of supersteps, is $\mathcal{O}($diameter$)$ for a given graph.
As a generalization, \emph{component-centric} models like Giraph++~\cite{giraph++}, Blogel~\cite{blogel} and our own GoFFish~\cite{goffish} have been proposed. Here, the unit of computation is coarser than a single vertex, and these mitigate the large communication overheads, lack of data co-location, and large number of iterations seen in Pregel, particularly for large-diameter graphs. 

While there has been a lot of literature on empirically evaluating these distributed graph processing platforms for different graphs and algorithms, they do not offer a formal basis of analyzing and generalizing these behaviors~\cite{lu2014large}. Graph applications are often irregular, and their runtime characteristics inherently hard to model. Further, these coarser models are sensitive to data locality and load imbalances caused by graph partitioning and placement~\cite{han2014experimental}. The large graph structure, distributed nature of partitions, and non-stationary behavior of graph algorithms makes them unwieldy to examine, and their runtime performance hard to predict.

In this paper, we propose a coarser version of the partitioned graph, that we term \emph{meta-graph}, as a analytical sketch over which to model and analyze the characteristics of distributed graph algorithms. Meta-graph are composed of meta-vertices that represent coarse components in the partitioned graph, such as weakly connected components (WCC), and meta-edges that indicate edges between vertices in the meta-vertices. In particular, we target coarse-grained component-centric models~\cite{blogel,goffish}, and draw a relative comparison of their behavior with vertex-centric models. Since partitioning is an intrinsic part of distribute graph processing, we take a two-step approach: First, we explore the impact of different partitioning strategies on the meta-graph for a graph. Next, we examine the behavior of graph algorithms for a given meta-graph that acts as a proxy for the original graph. This allows us to draw algorithmic inferences based on the coarse-sketch that is much smaller, without graph and partitioning specific artifacts. 

This work extends our previous work~\cite{dindokar:parlearning:2015} that was limited to examining the BFS algorithm for three smaller graphs using a single partitioning strategy. Here, we generalize that approach by formally introducing the meta-graph sketch, exploring three different partitioning strategies, include PageRank in the algorithmic analysis, and validate these results for 5 large graphs with spatial and powerlaw topologies. 

Our goal is to offer a methodological approach to analyze and gain insights on graph partitioning and algorithms, \emph{a priori} without having to implement complex partitioning and placement code, or run large experiments, for component-centric models. It is not our direct goal to propose new partitioning strategies to improve the performance. While we use our GoFFish~\cite{goffish} subgraph-centric model to empirically validate the analysis, this can be generalized to other component-centric frameworks such as Blogel~\cite{blogel} as well.

In this paper, we make the following contributions:
\begin{enumerate}[noitemsep,topsep=0pt,parsep=0pt,partopsep=0pt]
\item We introduce four \emph{common graph partitioning strategies} used by component-centric programming models, Hash, Default, Flat and Hierarchical~(\S~\ref{sec:partition}).
\item We present the idea of a \emph{meta-graph sketch} and offer a \emph{detailed analysis} of the impact of these partitioning strategies on the characteristics of the meta-graph~(\S~\ref{sec:metagraph}). We use two spatial and three powerlaw graphs to illustrate this analysis. 
\item We use the meta-graph for a partitioned graph to \emph{analyze the behavior} of two canonical graph algorithms, PageRank (PR) and Breadth First Search (BFS), designed using a subgraph-centric model, and contrast them against a vertex-centric model~(\S~\ref{sec:algos}). The analysis examines the supersteps, communication and computational complexity of the algorithms, and further correlates this analysis with the empirical results for these graphs and algorithms using GoFFish and Giraph.
\end{enumerate}

These contributions are complemented by \S~\ref{sec:related}, where we examine related work on component-centric graph algorithms, and analytical models for evaluating graph algorithms; and our conclusions presented in \S~\ref{sec:conclusion}.
%  \ysnote{Why is it important to solve this problem?}
% \ysnote{Why is it relevant to SC conference?}
% \ysnote{What is novel about the problem?}
% \ysnote{What is novel about the solution/analysis?}

\section{Related Work}%~~~\note{1.25 pages}}
\label{sec:related}

%We discuss related work on distributed graph processing platforms as background, empirical means to evaluate and analyze such graph platforms and algorithms, and formal methods to examine graph algorithms and other Big Data platforms. 

\subsection{Graph Processing Platforms} 
\emph{Google's Pregel} is an iterative vertex-centric programming model~\cite{pregel} %in which a user has to write the logic for a single vertex. 
based on  BSP execution model, which uses messages for state transfer between vertices across superstep boundaries. %where multiple workers on one or more machines independently operate on subsets of vertices in a graph, and execute the common user logic on all their vertices. Vertices are typically hashed onto workers to balance the number of vertices per worker. Messages are used for state transfer at superstep boundaries.
%The logic can compute on and update the vertex's value, process messages received, and emit messages to neighboring vertices. 
%A barrier synchronization ensures that messages generated by vertex computations are delivered in ``bulk'' to destination vertices, and then the next iteration of compute starts. 
The graph algorithm executes as a series of %these
 \emph{supersteps} till all vertices \emph{vote to halt} and have not generated new messages in a superstep. \emph{Apache Giraph}~\cite{giraph} is an open source implementation of Pregel.
%~\footnote{Apache Giraph. http://giraph.apache.org} is an open source implementation of Pregel. 
% ~\cite{bspvaliant}
\emph{Distributed Graphlab}~\cite{graphlab} contrasts with Pregel with asynchronous ``pull'' based state transfers between vertices %rather than barrier-synchronized push of messages at superstep boundaries. 
without the need for messaging. 
While potentially faster than Pregel, algorithms are harder to develop, analyze and the distributed locking protocols makes the implementation difficult.

Other component-centric platforms such as \emph{Giraph++}~\cite{giraph++}, \emph{GoFFish}~\cite{goffish} and \emph{Blogel}~\cite{blogel} have a coarse unit of computation than a vertex. While the former applies the user logic on a graph partition, the latter two offer a subgraph or a block %, or more generally, a WCC present in a partition 
 as the logical unit of computation. This allows coarse-grained computation, such as traversals on all vertices and edges present in the component, to take place in a single superstep, and bulk messaging between partitions happens at superstep boundaries. The graph is partitioned to increase the connectivity of vertices held by a worker, and reduce the edge-cuts across workers. Each component offers a degree of parallelism, and can be operated by an independent thread. This has the advantage of using shared memory graph algorithms in a subgraph, and distributed algorithms across supersteps. 

As a result, these frameworks can converge in fewer supersteps, and the messaging is limited to ones between subgraphs rather than fine-grained ones between vertices. However, while these may seem intuitive or be demonstrated experimentally, a formal analysis of the behavior of the component-centric programming models for different algorithms and graphs has not been undertaken as yet. We address this gap.

\subsection{Empirical Analysis and Optimization} 

Experimental observations have shown that workload imbalances in distributed graph-processing systems lead to slow convergence, high communication or computation cost. 
These imbalances are a function of the structural properties of graphs, diverse partitioning approaches and irregular runtime behavior of algorithms. \cite{salihoglu2014optimizing} offers algorithmic optimizations such as finishing computations serially, and on-demand edge cleaning to reduce communication cost and the number of supersteps. Others~\cite{yan2014pregel} describe desirable properties for distributed graph algorithms and proposed optimizations to some graph algorithms. System level optimizations such as ~\cite{shao2015page} including domain specific languages~\cite{hong2014simplifying} have been proposed to efficiently process large workloads. While a rich area for research, these optimizations have been proposed to address anecdotal limitations, and validate through experiments, rather than a formal analysis of the deficiencies and their resolution.

%Domain specific languages such Green Marl~\cite{hong2014simplifying} have been developed to ease of programming and offer optimized implementations for Pregel-like systems. System level optimizations are suggested in~\cite{shao2015page, zhou2014mocgraph} to efficiently process large message workloads. While a rich area for research, these optimizations have been proposed to address anecdotal limitations, and validate through experiments, rather than a formal analysis of the deficiencies and their resolution.

Efficient graph partitioning is integral to distributed graph processing systems %since it affects the execution for efficiency of jobs
for efficiency. 
%There has been significant work on this NP-hard problem~\cite{garey1974some}. 
The vertex/edge partitioning has been extensively researched~\cite{metis,scotch,bourse2014balanced}, to partition a graph into $k$-way vertex/edge-balanced partitions, but it does not consider the number of and sizes of components within each partition. %Edge centric partitioning~\cite{bourse2014balanced} has also been considered. %Literature shows that aggregation of  messages between hosts, like Pregel's combiner, the communication cost on an edge-balanced partitioned graph would be much lesser than a vertex-balanced one.
Formally understanding the impact of the partitioning on the graphs and graph algorithms is essential to select the appropriate partitioner, and should not require us to run the algorithm to understand the impact. Our proposed meta-graph offers an intermediate analytical sketch based on the partitioning of graphs to analyze their expected performance by different graph algorithms. We explore the impact of various partitioning algorithms of the meta-graph as well.
  
The \emph{a priori} partitioning strategies are complementary to runtime strategies that are tuned to specific graphs and algorithms based on performance monitoring. GPS~\cite{gps} and Pregel~\cite{pregel} automatically repartition vertices of the graph across the nodes of a cluster depending on their messaging patterns between supersteps. Runtime load monitoring for dynamic migration of vertices has also been explored in Mizan~\cite{mizan}. %Vertices from over-utilized workers are migrated to under-utilized ones.
 Our own work~\cite{ccgrid} describes elastic placement of partitions on cloud VMs, based on \emph{a priori} knowledge of non-stationary graph algorithms' behavior. While such dynamic strategies are beneficial to address minor deviations in the original partitioning, static analytical models can ensure that the initial partitioning itself is good to begin with for the given graph and algorithm. We address this problem.

\subsection{Formal Analysis of Big Data Platforms} 
There have been formal models developed to analyze the complexity and capabilities of Big Data programming models such as MapReduce~\cite{karloff2010model,sarma2013upper}. Such foundations help generalize the characteristics of the programming and data models, and offer stronger guarantees than just empirical results. Such methodologies are lacking for component-centric distributed graph processing abstractions, which we address here.

%Graph models have been developed~\cite{calvert1997modeling} to represent the topology of large networks such as Internet to study aspects of locality and hierarchy characteristics. 
%Graph traversal techniques such as BFS, random walks are used to sample large network topologies~\cite{najork2001breadth}. 
Authors in~\cite{fay2016predictive} have used theoretical analysis of BFS and shown that structural properties of the graph can be used to reduce computational bottlenecks of BFS. We take a similar approach to the analysis of distributed BFS and PR, using a meta-graph sketch as proxy for the partitioned graph.
% mislove2007measurement

Coarsening is a common technique used to reduce the size and complexity of the problem such that the reduced problem has similar characteristics to the original problem. It has been used in approximation algorithms for NP hard optimization problems~\cite{klein2010approximation}.  %such as Maximum-Weight Independent Set %Minimum-Weight Vertex Cover
In~\cite{chernoskutov2015heuristic}, coarsening is used for computing betweeness centrality in dynamically changing graphs. 
%Coarsening is used to create smaller representation of the graph on which changes are assessed and betweenness centrality is estimated for the original graph.
 Multi-level graph partitioning techniques also rely on coarsening for getting a good initial clustering ~\cite{metis}.  %to create smaller graphs which are structurally similar to the original graph
We leverage such an intuition in designing our coarse meta-graph sketch, and offer an analysis of the impact of partitioning on this sketch, and the sketch on graph algorithms.
% ,~\cite{safro2015advanced}
% Coarsening is also used in multigrid method to solve system of turbulent dynamic system. Mesh refining uses coarsening to reduce the dimensions of search space. ~\drnote{need to add citation for popular texts } 
% ~\drnote{should we add blockrank as an example for coarsening?} 
%The authors report significant speed-ups for PageRank and Distributed Minimum Spanning Tree algorithms on large graph datasets, even though they incur some overhead due to the cost of vertex migration. 
% Hierarchical Parallelization and Runtime Scheduling for Pregel-Like Graph Processing Systems
% http://ieeexplore.ieee.org/xpl/login.jsp?tp=&arnumber=7037707&url=http%3A%2F%2Fieeexplore.ieee.org%2Fiel7%2F7031670%2F7036227%2F07037707.pdf%3Farnumber%3D7037707
%Managing Large Graphs on Multi-Cores With Graph Awareness
%https://www.usenix.org/system/files/conference/atc12/atc12-final182.pdf
% Effective Techniques for Message Reduction and Load Balancing in Distributed Graph Computation
% http://arxiv.org/abs/1503.00626
% http://ieeexplore.ieee.org/xpl/login.jsp?tp=&arnumber=7004369&url=http%3A%2F%2Fieeexplore.ieee.org%2Fxpls%2Fabs_all.jsp%3Farnumber%3D7004369
% Wrangler: Predictable and Faster Jobs using Fewer Resources
% AMP Labs
\section{Component-centric Partitioning Strategies}%~~~\note{1 page}}
\label{sec:partition}
% \Note{Proposed mapping strategies are quite trivial. [repartition]}\\
% \Note{Focus is computational combinatorics? [Emphasize big data]} 
% \Note{sufficient system architectural considerations in designing the strategies [out of scope, Java ForkJoinPool]}
In a subgraph-centric framework, such as GoFFish,  the graph is partitioned onto distributed machines, and subgraphs within partitions identified as components for computation on those machines. The distribution of subgraphs, both in terms of their sizes and their numbers -- on each machine and across different machines -- has an impact on the load-balancing on the machines in each superstep. The barriered superstep causes less loaded machines to be idle while waiting for more loaded machines to complete processing, thereby reducing the average cluster utilization and increasing the makespan. Hence, one of the goals of the partitioning should be to balance the \emph{degree of component parallelism} in each machine while also balancing the \emph{size of each component}. %, the utilization of CPU cores and the time taken to complete a barriered superstep.  

%An extreme form of exposing parallel components and balancing their sizes is to use a vertex-centric model. So, 
To leverage the benefits of shared-memory processing that is available to coarse components, we should also try to \emph{maximize the size of each component}. One intuition that arises from this is to have as many large components as the number of processing cores in the cluster, ensure each component is as large as possible, and also balance their sizes -- assuming all cores are symmetric in their performance. This exposes as much coarse component parallelism as the number of processors, and provides the opportunity to process more of the graph in a few coarse supersteps. 

At the same time, the \emph{edge-cuts} between the partitions processed by workers should also be minimized. There are two factors here: the \emph{remote meta-edges} going between components on different machines, and the \emph{local meta-edges} between components on the same machine but owned by different workers/partitions. The former contributes to network latency in addition to the coordination overhead of spanning supersteps, while the latter is messaging between processes within the local host, but still requiring a barriered superstep for message exchange.

%when creating one partition per machine, which we term as \emph{Default Partitioning} below, offers a low CPU utilization of under $10\%$.

Partitioning large graphs is a hard problem, and there has been significant work on developing approximate solutions. Here, we leverage existing, well-understood partitioning algorithms that are applicable to component-centric framework, and whose analysis we will benefit from. 

%Since a single large subgraph, executing in a single core, dominates the execution time for a partition, creating as many partitions as the number of cores across the (symmetric) machines can help improve the utilization, but can exacerbate the communication costs or the number of supersteps taken to complete. We use this idea to propose two new partitioning strategies, besides the default, and analyze their consequences in the next sections. 

%\textbf{Preliminaries.} 
Formally, we consider the given graph that has to be partitioned as $\mathcal{G} = (V,E)$ where $|V| = n$ and $|E| = m$ are the number of vertices and edges, respectively. We are deploying the partitions over a cluster of $k$ symmetric machines, each having $c$ cores. %Let $\mathcal{M} = (\mathbb{V},\mathbb{E})$ be the meta-graph formed after the partitioning, where $\mathbb{V}$ is the total number of subgraphs (meta-vertices) created and $\mathbb{E}$ the number of meta-edges between them.

%Fig.~\ref{fig:part-strategy} illustrates the three partitioning strategies for a graph $G$ onto a cluster of $3$ machines having $2$ cores each. 

\begin{figure}[t]
\centering%~%
\includegraphics[width=\columnwidth]{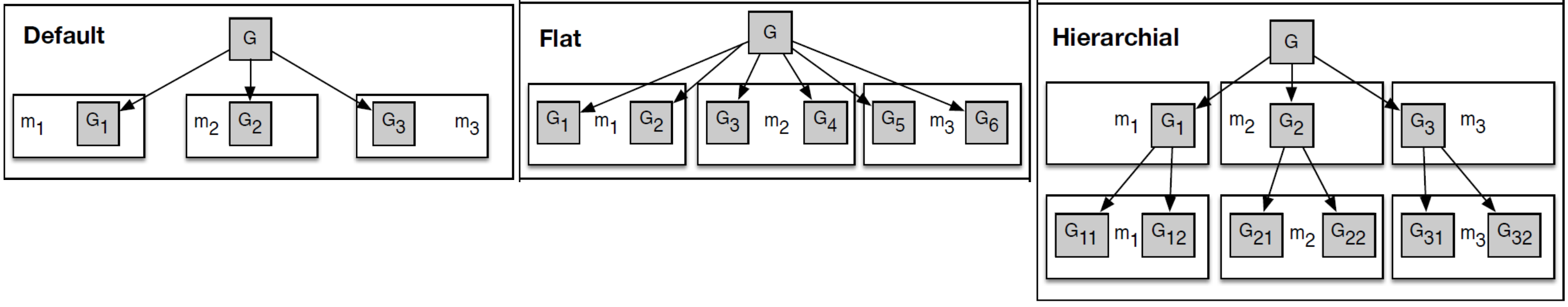}
\caption{Three partition strategies illustrated for a graph $G$, on a cluster of 3 machines having 2 cores each.}
\label{fig:strategy}
\vspace{-0.1in}
\end{figure}
\subsection{Default Partitioning [DP]} 

A na\"{i}ve partitioning approach is to divide the graph $\mathcal{G}$ into $k$ parts: $P_1 = (V_1,E_1),$ $P_2 = (V_2,E_2),$ $\ldots,$ $P_k = (V_k, E_k)$ using \emph{vertex-balanced partitioning} such that (1) $\sum_{i=1}^{k}|V_i| = |V|$ and $\forall i,$ $|V_i|~\approx~\frac{n}{k}$, and (2) the number of edge cuts between partitions is minimized. We distribute partition $P_i$ to machine $i$, and subsequently identify WCCs that lie wholly within $P_i$, which form the subgraphs. %For simplicity, we assume through the rest of the paper that the graphs are undirected, and thus, subgraphs are just connected components.
%~\footnote{For simplicity, we assume through the rest of the paper that the graphs are undirected, and thus, subgraphs are just connected components.} % We offer some thoughts in the conclusion on directed graphs.
%}.
 Existing algorithms such as METIS~\cite{metis} offer such vertex-balanced partitioning. Fig.~\ref{fig:strategy} illustrates this approach, and the two others.

A drawback of this strategy is that in practice, each partition ends up with one large WCC and hundreds of tiny subgraphs. Having one partition per machine causes $1$ core to be busy with the large subgraph while the remaining $c-1$ cores process the tiny subgraphs quickly in a superstep, and remain idle.% with unbalanced load across the cores of a multi-core machine.

\subsection{Hierarchical Partitioning [HP]} In this strategy, we initially partition $G$ into $k$ vertex-balanced parts, $P_1, \ldots, P_k$, that attempt to minimize edge-cuts between them, just like the DP strategy. Each partition is placed on its own machine, and further divided into $c$ vertex-balanced partitions, $P_{k,c}$, thus creating as many partitions as the number of cores. The connected components are identified within each of the $k \times c$ partitions to obtain the subgraphs. %As before, an algorithm like METIS can be used for both these levels.

HP generates as many partitions as the number of cores and if there is one large subgraph per partition, it will lower the load imbalance between the cores. Further, the first level of partitioning identifies $k$ partitions that are least connected to each other, and hence can lead to lower network transfer between supersteps. The second level of partitioning will, however, introduce intra-machine local meta-edges between the components in different partitions of a single machine.  These are analyzed in the next section. %and in the number of edge cuts across machine. However the total number of edge cuts across $k*c$ portion may increase.
 
\subsection{Flat Partitioning [FP]} In flat partitioning, we directly partition the graph $G$ into $k \times c$ vertex-balanced partitions. We distribute $c$ of the partitions to each of the $k$ machines, at random. The WCCs are identified within each partition in a machine to form subgraphs. 

This retains the benefit of HP by having as many partitions as the number of cores. However, since it does not consider the connectivity between partitions when placing them on different machines, the number of remote meta-edges and hence inter-machine communication costs may increase. At the same time, unlike HP, the total number of meta-edges, both remote and local together, would be fewer since the single-level partitioning has a global view of the whole graph. In FP as well as HP, the increase in the number of subgraphs, relative to DP, can have a negative impact on number of superstep. We empirically analyze these performance implications on different algorithms and graphs next.

\subsection{Hash Partitioning [HA]}
%For the sake of completeness, we also list hash partitioning which is used by vertex-centric models, but not by coarse grained models. 
Here, form $p=k \times c$ partitions, and each vertex is hashed based on its ID and placed in one of the $p$ partitions. This trivially ensures vertex balancing across the partitions and ensures vertex-parallel applications have uniform access to the CPU core. However, as we see later, these have poor communication complexity and are less suited for edge-intensive algorithms.

%%%%%%%%%%%%%%%%%%%%%%%%%%%
%% Analysis section
%%%%%%%%%%%%%%%%%%%%%%%%%%%
%\input{analysis.tex}  %% for article?

%%%%%%%%%%%%%%%%%%%%%%%%%%%
\section{Analysis of Partitioning Strategies' Impact on the Meta-Graph}%~~~\note{3 pages}}
\label{sec:metagraph}

We explore the impact of each partitioning strategy on the structure of the meta-graph, as this affects the analytical and empirical behavior of distributed graph algorithms that are discussed later. We introduce the meta-graph concept, and consider the number of vertices and edges, weights and the diameter of the meta-graph formed from a partitioned graph. %and their expected performance relative to vertex-centric versions of the algorithms.% different class of graph algorithms. 

\subsection{Definitions}
\label{sec:analysis:definition}
% We present formal definition of graphs, partitioning and meta graphs which will be used in subsequent section for analysis. 
%Let the undirected graph which is partitioned and on which computation is performed be 
Consider an undirected $G$ %$G = (V,E)$ where $|V| = n$ and  $|E| = m$.
 is $p$-way partitioned with $P = \{P_1, P_2, ...,P_p\}$ where $P_i = (V_i, E_i)$.  $V_i$ represents the vertices in partition $P_i$ and $ \bigcup_{i=1}^p V_i = V$  and $V_i \bigcap V_j = \varnothing$ $\forall i,j \in (1,p)$. Each partition $P_i$ has one or more the subgraphs, each of which is a connected component such that their vertices can be reached using local edges in that partition. As a result of partitioning, $G$ is decomposed into a set of $q$ subgraphs $SG = \{SG_1, SG_2, ..., SG_q\}$ across all partitions, where $q \ge p$. Let $SG_i = (V^s_i, E^s_i, R^s_i)$ where $V^s_i$ is the set of vertices in the connected component, $E^s_i$ is the set of local edges wholly within the connected component, and $R^s_i$ is the set of remote edges connecting the vertices in this subgraph with vertices in other subgraphs present in different partitions.

%  A vertex-balanced partitioning algorithm will also ensure $|V_i| \approx \frac{v}{k}$ while minimizing the number of edges crossing between partitions. 

We define a \emph{meta-graph} as $\widehat{G} = (\widehat{V} , \widehat{E})$ where each \emph{meta-vertex}, $\widehat{v}_i \in \widehat{V}$ represents a subgraph $SG_i$ with $q=|\widehat{V}|$, and a \emph{meta-edge}, $\widehat{e}_{jk}= (\widehat{v}_j, \widehat{v}_k) \in \widehat{E}$ represents the existence of remote edges connecting $SG_j$ to $SG_k$. A function, $weight_V[\widehat{v}_i]$ gives the number of vertices in $SG_i$ whereas $weight_E[\widehat{v}_i]$ gives the number of internal edges in that subgraph. Similarly, $weight[\widehat{e}_{jk}]$ gives the number of edges connecting vertices in $SG_j$ with vertices in $SG_k$. The sum of all the meta-vertex weights will equal $n=|V|=\sum_{i=1}^q weight[\widehat{v}_{i}]$, and the sum of all meta-edge weights will equal the edge cuts across partitions. %For e.g., see Fig.~\ref{fig:citp-metagraph}.
For our study, we assume a commodity cluster with $k$ symmetric machines each having $c$ CPU cores.

The meta-graph offers a useful tool to examine the performance behaviour of subgraph-centric graph algorithms on the partitioned graph. Next, we analyze the relative impact of the partitioning strategies on the meta-graph structure.

%Default partitiong , Flat Partitioning and Hierarchial Partitioning will be referred as DP, FP and HP respectively.
%\} where $P_i = (V^p_i, E^p_i)$
\subsection{Types of Graphs and Partitioning Setup}
\label{sec:graphs}
% For both our analysis and empirical results of different strategy 
%Graph partitioning is sensitive to the topology of the graph. 
We consider two classes of large-scale real-world graphs: \emph{spatial} networks, and graphs with \emph{powerlaw} distribution. Spatial graphs are characterized by a uniform degree distribution, a large diameter and a planar topology. Road networks and sensor networks fall under this category.
Powerlaw graphs feature a skewed degree distribution, a small diameter and a sparse structure. Their number of vertices with degree $d$ is given by $deg(d) = \alpha \times d^\beta$ where $\beta < 0$. Social networks and citation network fall in this class. These are shown in Fig.~\ref{fig:edgedeg}.

We use five undirected graphs, three powerlaw social networks, CITP~\footnote{http://snap.stanford.edu/data/cit-Patents.html}, LIVJ~\footnote{http://snap.stanford.edu/data/soc-LiveJournal1.html} and ORKT~\footnote{http://snap.stanford.edu/data/com-Orkut.html}, and two large diameter spatial networks, USRN~\footnote{http://www.dis.uniroma1.it/challenge9/download.shtml} and EURN~\footnote{http://www.cc.gatech.edu/dimacs10/archive/streets.shtml}, described in Table~\ref{tbl:graphs}. %Out of these, ORKT has the largest number of edges ~234M and smallest diameter of 9, and CITP has the smallest number of edges ~33M and has highest diameter of 22 among powerlaw graphs. In road networks, EURN has largest diameter $\approx15,740+$ and has almost twice the number of edges compared to USRN.  
We use METIS v4.0.1 with a default load factor of 1.03 for vertex-balanced partitioning used by DP, FP, and at both levels of HP. The sole exception is ORKT, which could not be partitioned by METIS (due to memory limitations), and instead we used Blogel's Vornoi partitioning~\cite{blogel} for the first (HP) or sole (DP, FP) level, and METIS for local partitioning within a machine (HP).

Table~\ref{tbl:graphs} summarizes the quality of the partitions generated by each strategy for each graph, for the 5-machine (40 CPU cores) and 10-machine (80 CPU cores) case. It lists the number of WCCs (or subgraphs, or meta-vertices) and meta-edges generated by each strategy, the fraction of vertices that are present in the large WCCs in each partition, the diameter of the meta-graph, and the fraction of edges that are cut.

\begin{table*}[t]
\centering
\scriptsize
\caption{Graph descriptions, and meta-graph details after partitioning on 5 and 10 Machines.} 
%\resizebox{!}{0.12\textwidth}{
\begin{tabular}{p{.8cm}|r|r||c||r|r|r|r|r|r||r|r|r|r|r|r}
\hline
{\textbf{Graph}} & $|V|~10^6$ & ${G}$ dia.$^1$ & \textbf{{Stra-}}& \multicolumn{6}{c||}{\textbf{5 Machines Meta-graph}} & \multicolumn{5}{c}{\textbf{10 Machines Meta-graph}}\\
\cline{5-16}
& $|E|~10^6$ & & \textbf{{tegy}} & Parts.$^2$ & $|\widehat{V}|^3$ &WCC\%$^4$& dia.$^1$&$|\widehat{E}|$ &Cut\%$^5$& Parts.$^2$ & $|\widehat{V}|^3$ &WCC\%$^4$& dia.$^1$&$|\widehat{E}|$ &Cut\%$^5$ \\
\hline
\hline
EURN & 50.91 & 15,740+ & DP &   5&  6&94\%& 2& 18&$\approx$ 0\%&  10& 13 & 100\%& 4& 44&$\approx$  0\%\\
     & 108.10        & & FP &  40& 49&95\%& 9& 199&$\approx$ 0\%&  80& 96& 99\%& 17&  422&$\approx$  0\%\\
                    && & HP &  40& 51&97\%&10& 196&$\approx$ 0\%&  80& 98& 99\%& 15&  428&$\approx$  0\%\\
\hline
USRN & 23.95 & 6,262 & DP &   5&  7& 93\%&  4& 16&$\approx$  0\%&  10& 11 & 100\%& 5& 36&  $\approx$  0\%\\
           & 58.33 & & FP &  40& 54& 96\%& 11& 216&$\approx$  0\%&  80& 96& 97\%& 4&  428&$\approx$  0\%\\
                  && & HP &  40& 49& 98\%& 11& 198&$\approx$  0\%&  80& 104& 95\%& 4& 452&$\approx$  0\%\\
\hline
\hline
ORKT & 3.07   & 9 & DP &   5&   5 & 100\% & 1& 20& 61\%&  10&  10 & 100\% & 1& 90& 72\%\\
     & 234.37 &   & FP &  40&  40 & 100\% & 1& 1,560& 83\%&  80&  81 & 99\% & 2& 6,471& 44\%\\
&&                & HP &  40& 108 &  99\% & 3& 1,898& 67\%&  80& 116 &  100\% & 3& 6,616&  78\%\\
\hline
LIVJ & 4.85 & 16 & DP &   5& 2,239& 99\%& 3& 796& 12\%&  10 & 2,251& 99\%& 3& 998& 16\%\\
       & 86.22 & & FP &  40& 3,153& 99\%& 4&  5,464& 29\%&  80 & 3,034& 99\%& 3& 10,854& 31\% \\
               &&& HP &  40& 6,564& 99\%& 4&  13,904& 28\%&  80 & 6,393& 99\%& 3& 19,410& 32\% \\
\hline 
CITP & 3.77 & 22 & DP &  5& 3,771& 99\%& 3& 376& 8\%&  10& 3,762& 99\%& 3& 522& 10\%\\
       & 33.04 & & FP &  40& 3,999& 99\%& 3& 3,572& 15\%&  80& 4,285& 99\%& 3& 11,060& 17\%\\
               &&& HP &  40& 4,488& 99\%& 3& 4,934& 16\%&  80& 4,306& 99\%& 3& 9,682& 18\%\\
\hline 
\end{tabular}
\\ $^1$ Diameter of the original graph or the meta-graph, as applicable.  \quad  $^2$ Number of partitions $p$ generated using the stategy. \quad  $^3$ Number of subgraphs $q$ present in the $p$ partitions.
\\$^4$ Fraction of vertices from the graph contained in the largest $p$ subgraphs.\qquad\qquad  $^5$ Fraction of edges in the graph that were cut by the partitioning.
\label{tbl:graphs}
\end{table*}

\subsection{Number and Weights of Meta-vertices} 
The number of meta-vertices represents the number of WCC (or subgraphs) present in all partitions. The partitioning strategies proposed have different effects on the number of meta-vertices for the two types of graphs considered.

Most real-world graphs contains at least one large WCC~\cite{faloutsos}. A vertex-balanced partitioning attempts to divide the vertices of this single large connected component equally among all partitions to have $|V_i| \approx \frac{|V|}{p}$ for all partitions $P_i$, while reducing edges between them. Due to the regular structure of \emph{spatial graphs}, the number of large subgraphs present in $q=|\widehat{V}|$ is close to the number of partitions $p$, one per partition, and this holds even for a large number of partitions. So, after DP performs a $k$-way partitioning onto $k$ machines, the resulting number of subgraphs is close to $k$. For HP and FP, we create $k \times c$ partitions and with one large subgraph per partition, we have $|\widehat{V}| \approx k \times c$ in both cases.
%This has a negative impact on CPU utilization since $k \times c$ cores are available. 

In Table~\ref{tbl:graphs}, we see that for the EURN and USRN spatial graphs, the number of WCCs is close to the number of partitions, for different partitioning strategies and different number of machines. For e.g., USRN has $7$ subgraphs when partitioned into $5$ parts, has $54$ subgraphs in $40$ partitions, and only increases to $96$ subgraphs with $80$ partitions on $10$ machines. A similar trend is observed for EURN too, with the number of WCCs almost the same as the number of partitions, such as $6$ WCCs from $5$ partitions for DP on 5 machines, or marginally higher, with $98$ WCCs for $80$ partitions using HP on 10 machines. This shows a slow growth in number of subgraphs $q$ that remain close to the number of partitions $p$. 

We also see that while the number of WCCs in a partition is close to $1$, the largest $p$ WCCs only includes $\sim96\%$ of the vertices in the graph, on average, with $4\%$ of vertices present in the remaining $(q-p)$ subgraphs. This means that the few non-dominant subgraphs are however not trivial in size, and may contain $100,000$'s of vertices -- such as $3,131,641$ for EURN on 5M using DP.

\begin{figure*}[t!]
\vspace{-0.1in}
\centering%~%
  \subfloat[\emph{Powerlaw Graphs} (Log scale for X Axis)]{\includegraphics[width=0.45\textwidth]{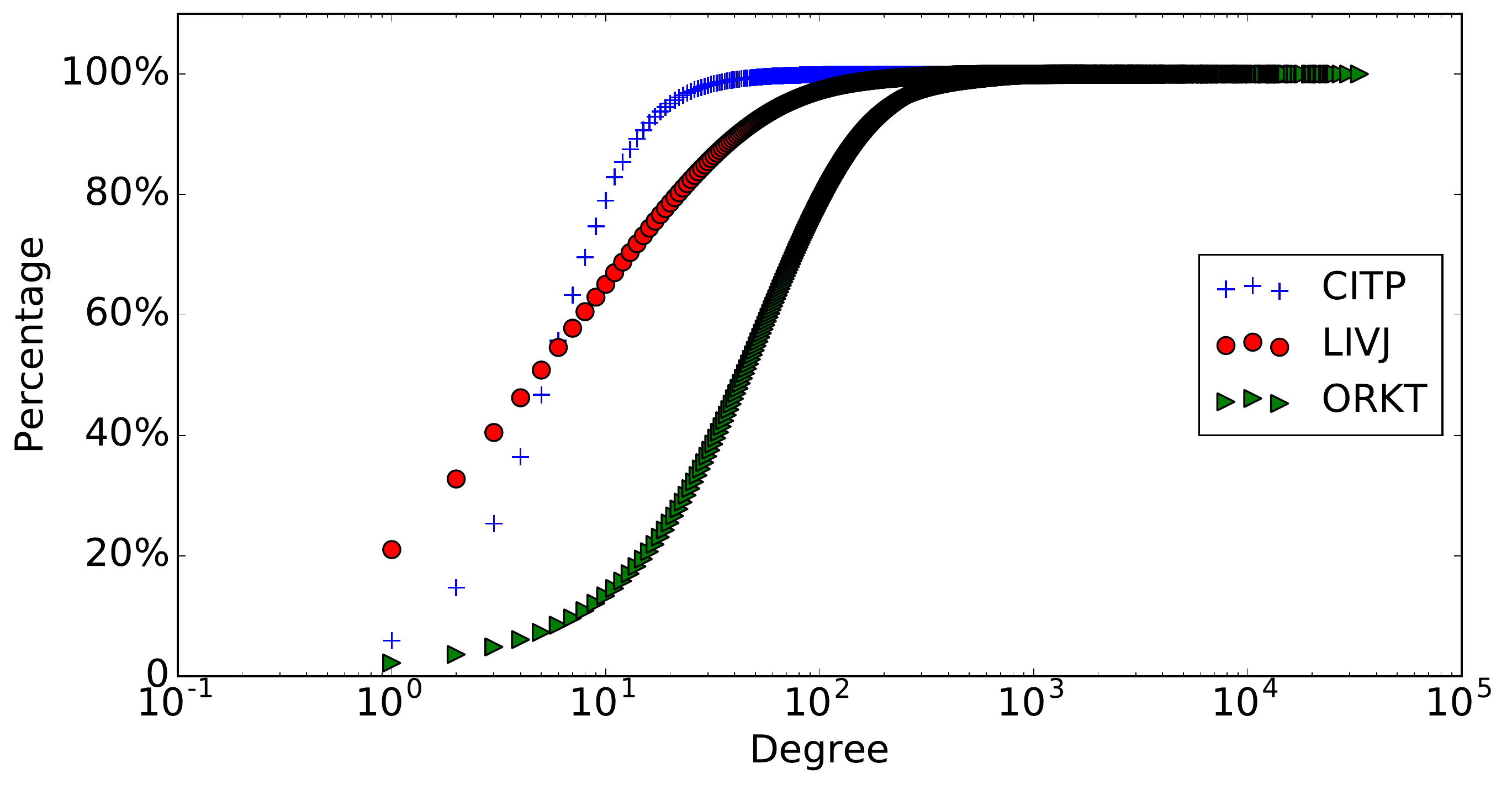}
    \label{fig:edgedeg:power}
  }
  \subfloat[\emph{Spatial Graphs}]{\includegraphics[width=0.45\textwidth]{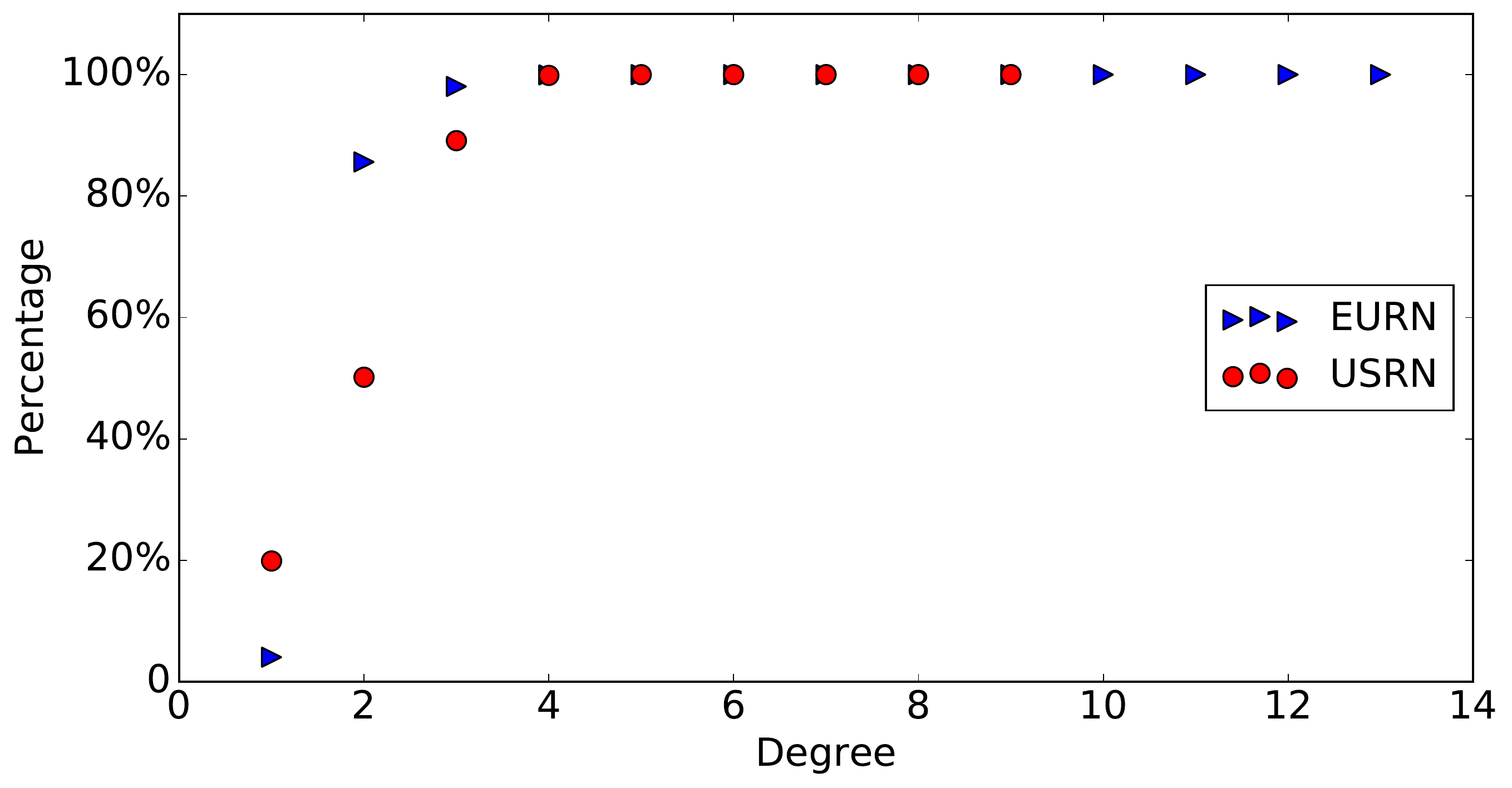}
    \label{fig:edgedeg:spatial}
  }
\caption{Cumulative distribution function (CDF) of vertex frequencies for different edge degrees, in the $5$ graphs.}
\label{fig:edgedeg}
\vspace{-0.1in}
\end{figure*}

For \emph{powerlaw graphs}, $|\widehat{V}|$ has a more subtle relationship as the number of partitions increase.
While applying FP for powerlaw graphs, the vertex-balanced partitioning tries to place the top $k \times c$ vertices with the highest edge degrees into different partitions to reduce edge cuts between them. 

On the other hand, for HP, in the first level of partitioning, vertex-balancing places the top $k$ vertices with the highest edge degrees into $k$ different partitions. Often, graphs like LIVJ and CITP have vertices that are densely connected components with a star topology, i.e., a single vertex with high edge degree connected to many vertices with small edge degrees, as shown in Fig.~\ref{fig:edgedeg:power} where these two graphs primary have vertices with degrees smaller than $10$. 

When performing the second level of vertex-balanced partitioning within each machine, the subgraph containing the star-topology vertex may need to be split to ensure vertex balancing.  This can result in partitions containing thousands of singleton or tiny subgraphs which are connected by remote edges to a subgraph containing the high edge-degree vertex present in another partition in the same machine ~\cite{lim2015discovering}. So, for HP, the number of meta-vertices $|\widehat{V}|$ will increase much more rapidly than FP, relative to DP and also as the number of machines or cores increases. Further, the weights of the meta-vertices will also have a skewed distribution as it can contain many meta-vertices with very small weights.

On the other hand, for powerlaw graphs like ORKT which have a \emph{small-world behavior}, hence a small diameter, the first level of partitioning will result in a densely connected subgraphs within each partitions. As a result, when they are partitioned further in the second level, they do not result in many more smaller subgraphs, but do end up having a higher edge cut\% than powerlaw graphs with larger diameters like LIVJ and CITP  ($16$ and $22$, respectively). This is evident in Fig.~\ref{fig:edgedeg:power}, where ORKT has a large fraction of vertices with edge degree of $75$ or more.

% \ysnote{For ORKT, there is more of a clique model due to a small diameter of $9$, so the number of unit/small SG is few/none. But CITP and LIVJ may have long chaings/hub+spoke due to larger dia of $16+$ which cause many SG's. Looking at the edge degree distribution of G and MG can help confirm this.}\ysnote{For the MG, for ORKT, its a fully connected graph, similar to original graph. For LIVJ and CITP, the MG has a powerlaw distribution, also similar to original graph.}
% ~\drnote{should we distinguish ORKT as small world graph with dense nature from LIVJ and CITP having sparse nature? It does not follow as per above description of HP }

In Table~\ref{tbl:graphs}, the three powerlaw graphs ORKT, LIVJ and CITP further confirm these behaviors. The number of WCCs for 5 partitions (on 5 machines) using DP is smaller than for 40 partitions (on 5 machines) using FP for all these graphs. While for ORKT, the number of subgraphs equals the number of partitions, at 5 and 40 respectively, due to its small-world nature, for LIVJ and CITP, the number of subgraphs is much larger than the number of partitions due to their hub-and-spoke model with larger diameters. We also see that the number of WCCs for HP are significantly higher than for FP for the same number of partitions ($q\gg p$), on both 5 and 10 machines, as we expect based on the above analysis. In fact, for LIVJ, HP gives twice the number of subgraphs as HP. 

We also see that the largest subgraph in each partition has 99\% of all vertices in the partition, for LIVJ and CITP, further validating that a large number of tiny subgraphs totaling up to 1\% of vertices in the partition are present. For ORKT, however, these large subgraphs have 100\% of all vertices in most cases, thus limiting the number of smaller subgraphs.

% \ysnote{not sure of the latter since vertex balancing should take care of this?}
% \ysnote{as there may not be $c$ different vertices with a high degree that can be placed in the $c$ partitions. Not clear why FP can have $k*c$ large degree vertices but HP cant find them. Need to describe that HP's k partitions may ``pull'' vertices k++ asymmetrically}. 

\subsection{Number and Weights of Meta-edges}
% \subsection{Relation between Partitions and Edge Cuts}
% \label{sec:part-edge-cuts}
%\ncnote{maybe move to section 4}
Let $\chi(G, p) = \sum weight[\widehat{e}_{ij}],$ $\forall \widehat{e}_{ij}\in \widehat{E}$, be the \emph{number of edge cuts} for a graph $G$ partitioned across $p$ partitions, given as the sum of its meta-edge weights. 

Minimizing $\chi(G,p)$ for a given $p$ is an NP-complete problem~\cite{garey1974some}. There exist many heuristics based on multilevel partitioning~\cite{multilevel-part,metis} and spectral partitioning~\cite{spectral-part}. For our analysis, we use METIS~\cite{metis} to perform vertex-balanced partitioning. We treat it as black box and assume that it gives a near-optimal value of $\chi(G,p)$.
Donath, et al~\cite{donath} give a lower bound on the minimum number of edge cuts as:
\[ \chi(G,p) \geq \frac{n}{p} \times \sum_{i=1}^p(\lambda_i), \] where $n=|V|$ and $\lambda_i$ is the $i^{th}$ largest eigenvalue of $L(G)$ where $L(G)$ is the graph laplacian given by $L(G) = deg(G) - adj(G)$, where $deg(G)$ is the degree matrix (a diagonal matrix with degree of vertices) and $adj(G)$ is the adjacency matrix.

For \emph{spatial planar graphs}, due to an even edge degree distribution (Fig.~\ref{fig:edgedeg:spatial}) and a planar topology, $\chi(G,p)$ increases linearly with $p$. Also, as spatial graphs are easy to partition optimally, the number of edge cuts are much lower than in the case of powerlaw graphs. 
For \emph{powerlaw graphs}, their eigenvalues also display a powerlaw distribution~\cite{eigenpower}. As a result, $\chi(G,p)$ initially increases rapidly as the number of partitions increase from $p=1$, but for larger values of $p$ this growth slows following an exponential decay. 

% As we observed in \S~\ref{sec:part-edge-cuts}, in the case of powerlaw graphs $\chi(G, p)$ increases rapidly with the increase in the number of partitions, for small values of $p$, but the rate of growth decays for larger values of $p$. For spatial graphs, $\chi(G, p)$  varies linearly with $p$. 
When we consider $k$ partitions for DP with $\chi(G, k)$ meta-edge weights, both HP and FP, which have a higher number of $k \times c$ partitions, will have a similar relative increase in their meta-edge weights, $\chi(G, k \times c)$, based on the type of graph. % relative to Increase in $\sum(weight[\widehat{e}_{ij}]) $ $\forall \widehat{e}_{ij}\in \widehat{E}$ for both HP and FP will follow similar trend. 
However, if the vertex-balanced partitioning algorithm offers near-optimal results, the growth in edge cuts, and hence sum of meta-edge weights, should increase more rapidly for HP than FP relative to DP or as the number of machines increases. The intuition is the same as for meta-vertices. For HP, we identify the $k$ partitions with tightly connected vertices and then try to further partition each into $c$ parts, causing deeper edge cuts. In FP, we identify the $k \times c$ partitions with minimal connections between each in a single pass. This effect will be more acute for powerlaw graphs with dense edge degrees. %observation will be more apparently visible in case of graphs having power law distribution as they are more difficult to partition.

We observe these trends in Table~\ref{tbl:graphs}. For the spatial graphs USRN and EURN, the edge cut fraction is always close to $\approx 0\%$, exhibiting a near ideal partitioning with minimal edge cuts. For the powerlaw graphs, we consider 5, 10, 40 and 80 partitions created by DP and FP on 5 and 10 machines. There is a gradual rise in the edge cut \% from 5 to 40 partitions, while it plateaus out between 40 to 80 partitions, in fact reducing in percentage terms for ORKT from $83\%$ to $44\%$, indicating that the absolute number of edge cuts have not gone up much. HP in general has similar or more edge cut \% than FP, with ORKT on 5 machines being the only exception where FP has an uncommonly high $83\%$ edge cuts.%\ysnote{FIXME. Why? whats going on with this? This seems an outlier even between 5 and 10 VMs for FP}

%\ysnote{Do we see this in the metagraph in the Table? How is the change in edge cuts between large powerlaw like ORKT and smaller ones like CITP when going from 5 to 10 machines/5p, 10p, 40p, 80p?}

%Number of meta  edges will vary also on number of meta vertices and follow some of the trends as described in previous section.

\subsection{Structure and Diameter of Meta-graph}
Meta-graphs for both powerlaw and spatial graphs exhibit a \emph{recursive behaviour}, whereby the structure of the meta-graph resembles the original graph. This indicate their suitability as a coarse-grained sketch for analysis. For \emph{spatial graph}, meta-vertices represent spatially proximate connected components and retain a uniform degree distribution and a planar topology.

In case of \emph{powerlaw graphs}, %with \emph{powerlaw distribution},
 as we increase the number of partitions $p$, we have a few meta-vertices having large edge degrees representing subgraphs containing the high degree vertices of the original graph. These meta-vertices are in turn connected to many meta-vertices representing small subgraphs present in other partitions which, as we have discussed, the partitioning algorithm has placed remotely to balance the number of vertices in each partition. Moreover we expect the degree distribution to be more skewed in case HP. % due to the imbalance in number of vertices having larger degree across partitions due to multilevel partition.  % disconnected from the vertex with the large degree,
% \ysnote{Can we get the edge degree distribution for each subgraph to see if this is true for spatial and planar graphs? How about the meta-edge degree distribution for the meta-vertices?}
This recursive behavior of degree distribution is empirically confirmed for both power-law and spatial graphs. For ORKT, this extends to the meta-graph being a complete graph, being a consequence of its dense degree distribution. % connected graph with average vertex degree is 76. }

%The \emph{diameter} of a graph is the longest shortest path between any two pairs of vertices. 
The diameter of the meta-graph $\widehat{G}$, $d(\widehat{G})$, for DP must be less than or equal to the diameter of the original graph $G$. The proof for this is straight forward. If $d(\widehat{G}) > d(G)$ then the path representing the diameter in $\widehat{G}$ can be expanded into a larger path in $G$ by expanding each meta-vertex $\widehat{v_i}$ into the subgraph it represents. As each subgraph contains at least one vertex, we get a path which is longer than the diameter. Hence, by contradiction, we prove $d(\widehat{G})  \leq d(G)$. % \ysnote{Write the previous proof a bit more clearer}
We can similarly prove that the diameter of meta-graph from DP is $\leq$ diameter of meta-graph from HP since meta-vertices in HP are created by splitting meta-vertices in DP. However, the diameter of the meta-graph from FP does not hold such a strict relationship.

% \subsection{Comparison with Vertex-centric Platforms}
% \ysnote{talk about how Giraph uses hash partitioning, the number of ``meta-vertices'' is same as number of vertices in Giraph, the expected number of local and remote edge cuts, balancing of vertices between partitions, diameter being same as whole graph. Some of this may be present in background sections.}
% %  which hashes a vertex's ID to one of the available machines to balance out the 

\section{Analysis of Algorithms using Meta-graph}%~~~\note{3 pages}}
\label{sec:algos}
%The above discussions offer a valuable foundational methodology which can be extended to other partitioning approaches as well, and their impact on the meta-graph. 
%Next, we show how given a meta-graph for a partitioned graph, we can study their properties to analyze the behavior of subgraph-centric graph algorithms. We also compare the expected behavior of the subgraph-centric algorithm with a vertex-centric one.
In this section, we show how given a meta-graph for a partitioned graph, we can study their properties to analyze the behavior of subgraph-centric BFS and PR algorithms. We also compare the expected behavior of the subgraph-centric algorithm with a vertex-centric one.

%The overall execution time (or makespan) for the algorithm in itself depends on various factors for a component-centric graph algorithm. Specifically, the it relies on the size and connectivity of each component, the computational cost to process each component by the algorithm, the communication cost between components across supersteps, the number of supersteps, and the skew between the wall-clock times taken by different partitions. We address these aspects for PR and BFS algorithms below.

\subsection{Setup for Algorithms}

We corroborate the analysis below with empirical results too. We run the PR and BFS algorithms using several real-world graphs, with our \emph{GoFFish} subgraph-centric platform~\cite{goffish}, and with \emph{Apache Giraph}~\cite{giraph} as a baseline vertex-centric platform based on Google's Pregel. We use the three powerlaw social networks graphs (ORKT, LIVJ, CITP) and two large diameter spatial networks (USRN, EURN) from Sec.~\ref{sec:graphs}, partitioned using DP, FP and HP for GoFFish, and for Giraph, using its default hash partition (HA).

 We run our experiments on a $24$-node commodity cluster with each node having one AMD Opteron 3380 (8~cores, 2.6GHz) CPU, 32~GB RAM and 256~GB SSD connected by Gigabit Ethernet. We use CentOS 7, and Giraph v1.1, Hadoop/Yarn v2.6 and GoFFish v2.6 run on JDK v7. GoFFish was modified to support the different partitioning strategies. %For comparison we use Giraph v1.1 running on YARN v2.2, and we assign each container one core and 4GB memory. 

We run BFS and PR algorithms using GoFFish and with Giraph on the above graphs. Specifically, we run a Single Source Shortest Path (SSSP) algorithm from a source vertex for a undirected graph with all edge weights 1, which is effectively a BFS. All experiments we run thrice, using different sources vertices in case of BFS, and, the averages reported.

%\begin{figure*}[t!]
%\centering%~%
%  \subfloat[5 Nodes]
%  {\includegraphics[width=0.33\textwidth]
%  {CDF_CITP}
%    \label{fig:Message-SSSP-5M}
%  }
%  \subfloat[10 Nodes]{\includegraphics[width=0.50\textwidth]{Message-SSSP-10M.jpg}
%    \label{fig:Message-SSSP-10M}
%  }
%\caption{MESSAGES FOR SSSP}
%\label{fig:CDF-Powerlaw}
%\end{figure*}
%

\subsection{PageRank (PR)}
%vertex centric computation $\mathcal{O}(n/p)$, 
PageRank is an iterative algorithm that % , and each iteration consists of three phases for each vertex: (1) Calculate the $sum$ of $PR$'s from neighboring vertices, (2) Update the local vertex's $PR$ by dividing the $sum$ by the out-degree, and (3) Send updated the $PR$ to neighboring vertices. The PR algorithm
runs for a fixed number of $30$ iterations (supersteps) for both subgraph- and vertex-centric models~\cite{pregel, goffish}. The behavior of each superstep is identical in terms of time complexity, so we can trivially extrapolate from a single superstep to several. 

A sequential/shared-memory version of PR has a computation complexity of $\mathcal{O}_c(|V| + |E|)$, since for every iteration, a total of $|E|$ additions and $|V|$ divisions is performed; and a communication complexity of $\mathcal{O}_m(|E|)$, as messages generated to every neighbor per superstep. Next, we discuss their behavior for the distributed formulations.

\begin{figure*}[t!]
\centering%~%
  \subfloat[\emph{5 Nodes}  (EURN values trimmed for display)]
  {\includegraphics[width=0.4\textwidth]{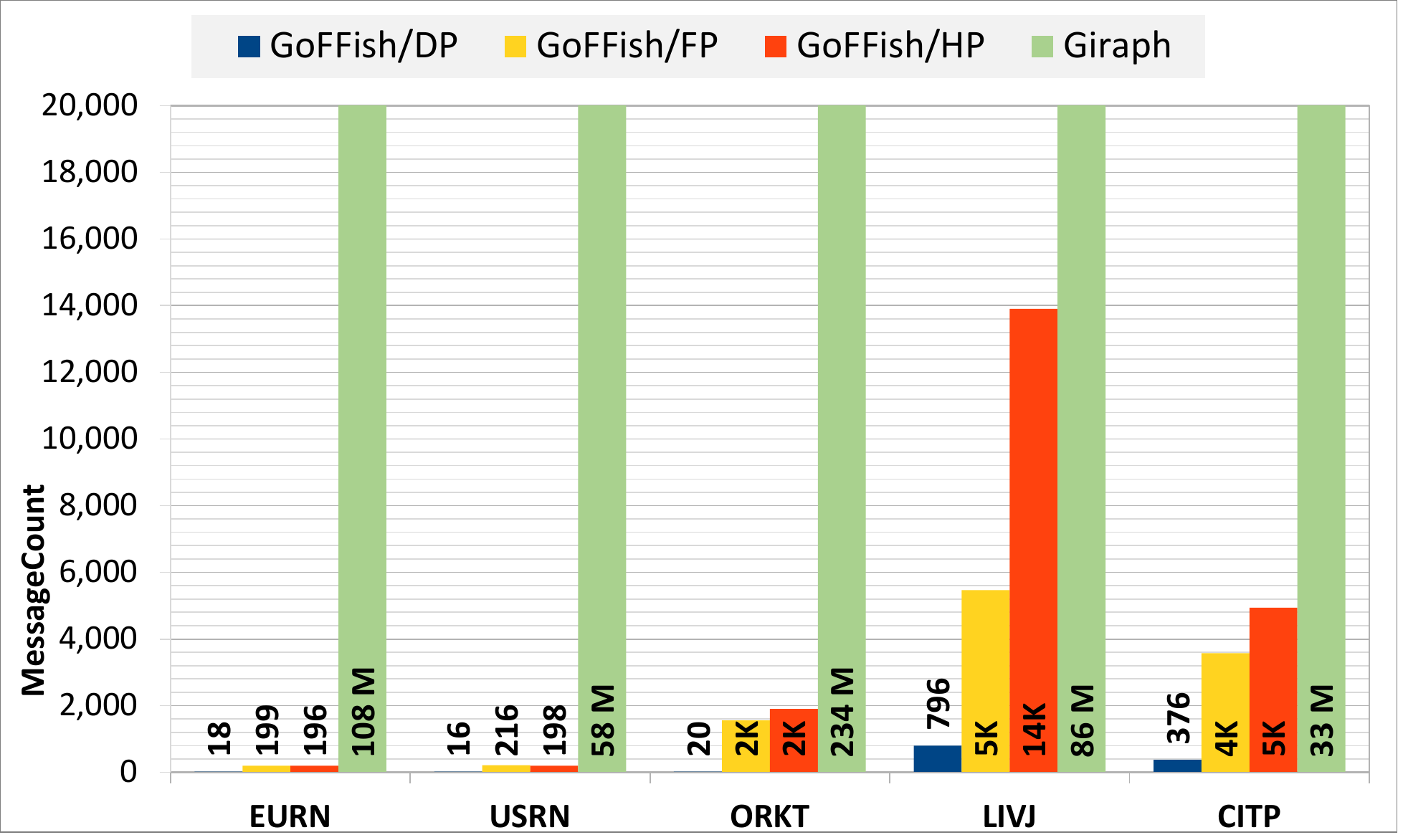}
    \label{fig:Message-PR-5M}
  }~~~~
  \subfloat[\emph{10 Nodes} (EURN values trimmed for display)]{\includegraphics[width=0.4\textwidth]{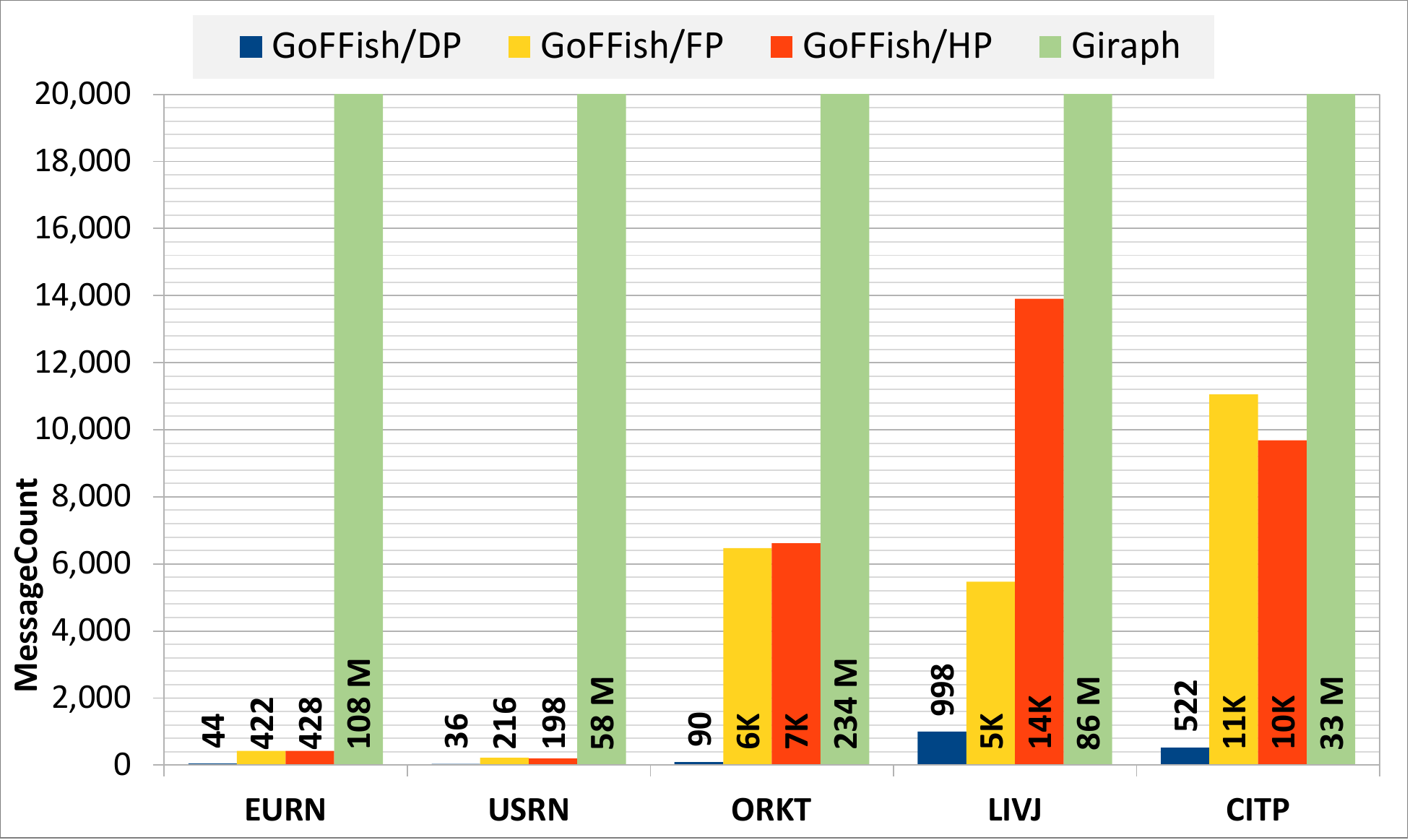}
    \label{fig:Message-PR-10M}
  }
\caption{Number of messages exchanged between components \emph{per superstep} during the PR.}
\label{fig:Messages-PR}
\vspace{-0.1in}
\end{figure*}

\subsubsection{Communication Complexity}
We distinguish between communication costs for in-memory message transfer for vertices (or subgraphs) co-located on the same machine (\emph{local messages}), and \emph{remote message} transfer over network between vertices (subgraphs) on different machines; the latter is more relevant and we emphasize that. 

For a \emph{vertex-centric model}, the expected number of remote edges in a hash partition of $p$ partitions for a graph $G$ is given by $\mathbb{E}^H = (1-\frac{1}{p}) \cdot |E|$, and conversely, the expected number of local messages is $\frac{|E|}{p})$~\cite{powergraph}. For larger values of $p$, $\mathbb{E}^H \approx |E|$. So, the remote communication cost, which happens in parallel across $p$ cores, is given by $\mathcal{O}_m^r(\frac{|E|}{p})$ for large values of $p$.

When we consider the messages \emph{per superstep} for running PR on Giraph (Fig.~\ref{fig:Messages-PR}), we see the total number of messages (both local and remote) is exactly equal to the number of edges in each graph, for all $5$ graphs on both $5$ and $10$ machine. E.g., from Table~\ref{tbl:graphs}, we have $108~M$ messages for EURN, for $5$ and $10$ machines, and $33~M$ messages for CITP.

In the \emph{subgraph-centric approach}, subgraphs contain both \emph{internal edges} within vertices in the subgraph, and remote edges to subgraphs in other partitions. % ~\footnote{By definition, there can be no vertices from subgraphs in one partition to subgraphs in the same partition.}.
Of these remote edges, for FP and HP, some could be incident on partitions in the same machine, while others in partitions in different machines, while for DP, by definition, all remote edges are incident on partitions in different machines.

In all three partitioning approaches, the total number of remote messages sent between subgraphs in different partitions in one superstep for PR will be equal to the sum of their meta-edge weights, which is also the number of remote edge, $\sum_{i=1}^q weight[\widehat{e}_{jk}]$. In platforms such as GoFFish, where logical messages between the same pair of subgraphs are batched together into a \emph{single physical-message}, we will find that the number of physical messages is the number of meta-edges $|\widehat{E}|$. So, the remote physical-message communication complexity for each superstep of PR using DP, FP and HP, which happens in parallel across $p$ partitions, is given by $\mathcal{O}_m^r(\frac{|\widehat{E}|}{p})$.

However, in case of HP and FP, $c$ partitions remain inside a single machine. As result, the real network bandwidth used per superstep depends on the number of edge cuts between the machines. For HP the number of edge cuts across machines is the same as DP, but FP will have a higher number of inter-machine edge cuts as $k \times c$ partitions are randomly placed on $k$ machines, with $c$ partitions each.

From Table~\ref{tbl:graphs} and Fig.~\ref{fig:Messages-PR}, we see that the number of physical messages exchanged between subgraphs in each superstep for the partitioned graphs is identical to the number of meta-edges in their respective meta-graphs. For e.g, we see that ORKT on $5$ machines passes $20$, $1560$ and $1898$ messages for its DP, FP and HP partitioning, which is identical to the number of meta-edges that each of their meta-graphs have in Table~\ref{tbl:graphs}.

% \ysnote{FIXME: It would be nice to have the number of logical messages sent between messages, not just physical}

% For graph $G$ that is partitioned onto $k$ machines with $c$ cores each using HP and FP, let $\chi^{HP}(G,k\times c)$ and  $\chi^{FP}(G,k \times c)$ denote the number of edge cuts % \ysnote{is this edge cuts between partitions or edge cuts between partitions in different machines?} \ncnote{across partition}
% , respectively. FP use the standard vertex-balanced partitioning to divide the graph, and hence, $\chi^{FP}(G,k \times c) = \chi(G,k \times c)$. In case of HP, applying the partitioning  due to application of partition algorithm twice we get $\chi^{HP}(G,k \times c)  > \chi^{FP}(G,k \times c)$ as described  in  (Sec~\ref{sec:part-edge-cuts})\ysnote{explain, referring to metagraph section earlier, if necessary}.

\begin{figure*}[t!]
\centering%~%
 \subfloat[\emph{PR on Giraph}]
 {\includegraphics[width=0.4\textwidth]{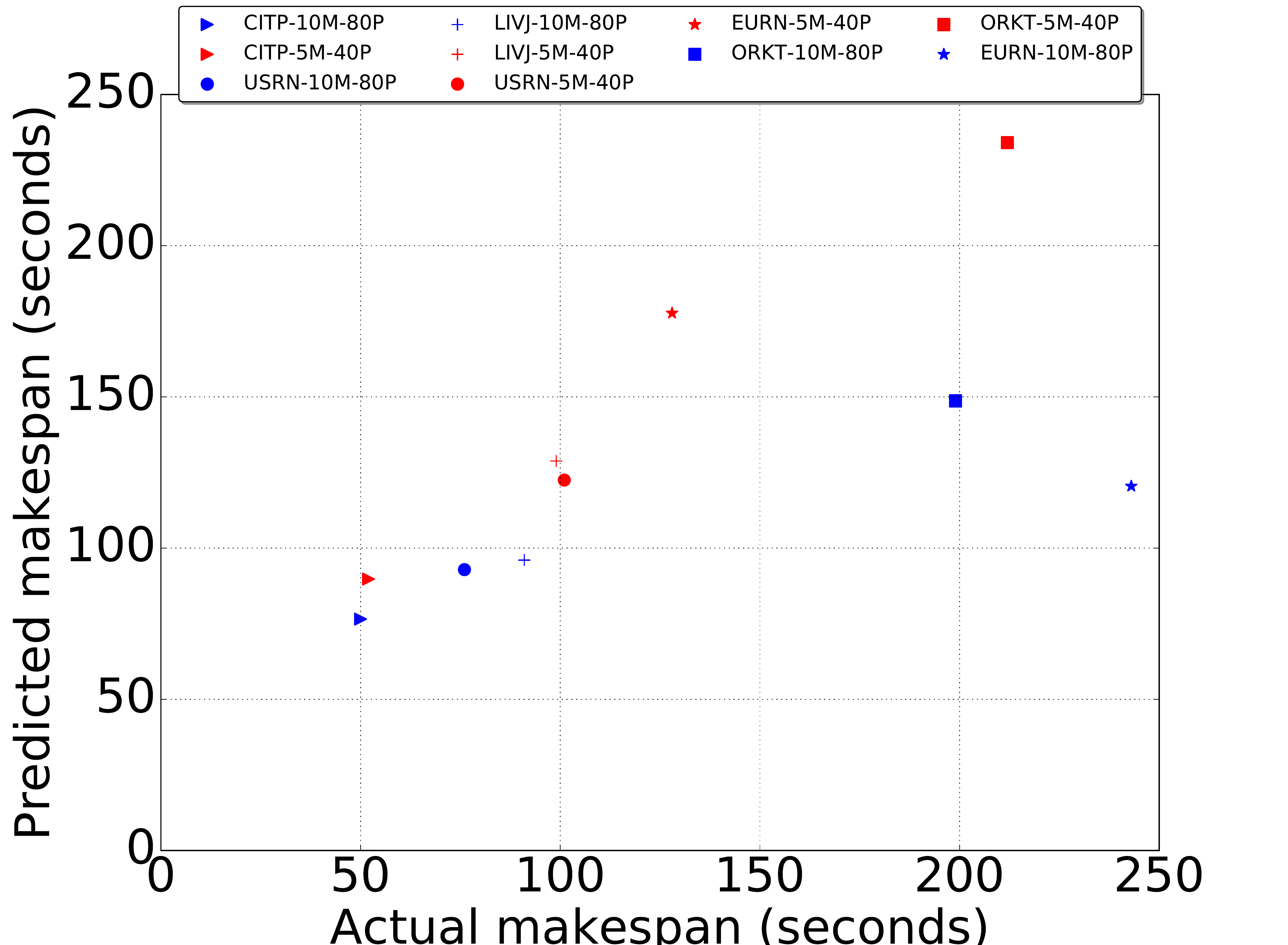}
   \label{fig:pr:makespan:giraph:scatter}
 }
 \subfloat[\emph{PR on GoFFish}, using Meta-graph Analysis]{\includegraphics[width=0.4\textwidth]{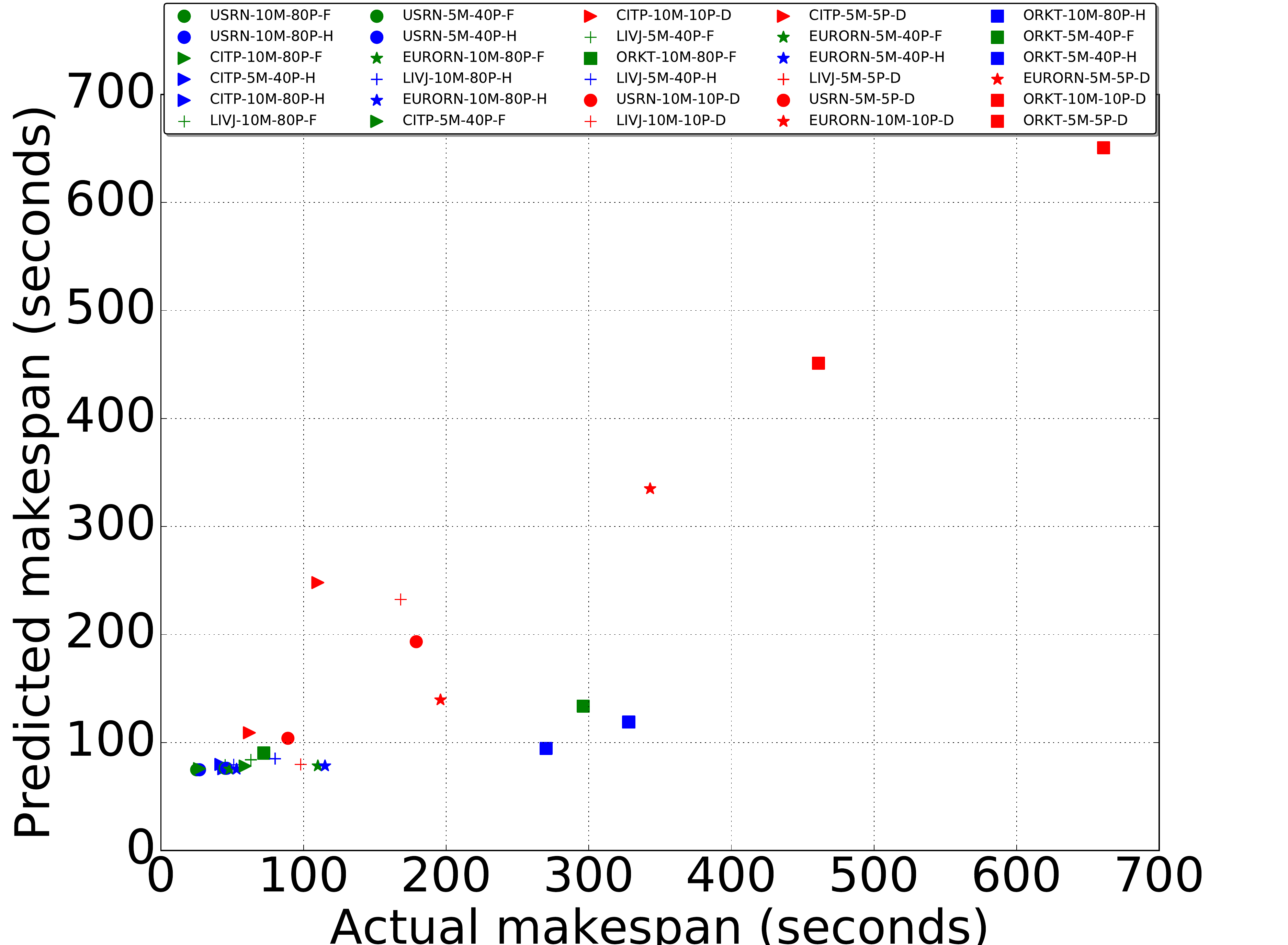}
   \label{fig:pr:makespan:goffish:scatter}
 }
\caption{Scatter plot between observed and expected makespan times for PageRank based on analysis.}
\label{fig:pr:makespan}
\vspace{-0.1in}
\end{figure*}

\subsubsection{Computational Complexity}

For a \emph{vertex-centric} framework using hash partition, the number of vertices $|V|$ will be equally distributed per partition/core $c$, since $|V| \gg c$. So when running PR on $p=k\times c$ partitions, where $k$ is the number of machines each having $c$ cores, each core operates on $\frac{|V|}{p}$ vertices. Ideally, if the edges are also equally distributed among the partitions, the computational complexity per superstep is $\mathcal{O}_c(\frac{|V|+|E|}{p})$ % and communication cost is $\mathcal{O}_c(\frac{|E|}{k \times c})$
, and uniform for each core. 

This holds true for spatial graphs, but for powerlaw graphs, the edge degree skew can impact the complexity and cause it to be unbalanced across cores, i.e., if the difference in edge degrees between the vertex with the highest degree and the vertex with the $(k \times c)^{th}$ degree is significant, we can expect the computation costs to be unbalanced and hence cause non-uniform utilization of the CPU cores.

% \note{internal edges in the largest subgraph for a single partition is $\approx \frac{|E|}{k}$.} \ysnote{this moves to computational complexity}

In a \emph{subgraph-centric model}, the computational complexity for performing PR on one subgraph depends on the internal vertices and edges for that component as well as its remote edges, since they all contribute to the PR update calculation. Hence, we have the computational complexity for a subgraph $SG_i = (V^s_i, E^s_i, R^s_i)$ given by $\mathcal{O}_c(|V^s_i|+|E^s_i|+|R^s_i|)$.

The computational cost for a partition is based on the largest subgraph in that partition, provided one subgraph per partition dominates, and that for a superstep depends on the largest subgraph across all partitions. Thus, when from the meta-graph, we have the computational complexity for a superstep based on its largest meta-vertex as  $\mathcal{O}_c\big(\max_{\widehat{v_i} \in \widehat{V}, \widehat{e_{jk}} \in \widehat{E}~\forall~(\widehat{V}, \widehat{E})\in \widehat{G} }$ $(weight_V[v_i] + weight_E[v_i] + weight[e_{jk}])\big)$.

%\ysnote{Relate giraph (V+E) to its makespan, goffish to largest SG using scatter plots}
Fig.~\ref{fig:pr:makespan:giraph:scatter} shows a scatter plot between the expected computational time, based on our analysis, and the observed makespan for running PR using Giraph. We see a close correlation between the two, with the outliers being seen only for the large graphs EURN and ORKT, on $80$ partitions.

% per superstep is determined by the subgraph taking the most computation time. For DP, the size of the largest subgraph, $\max(weight[\widehat{v_i}] )$ $\forall \widehat{v_i} \in \widehat{V}$ is approximately $\frac{|V|}{k}$. 
% \ysnote{how about utilization}
% Assuming a single subgraph dominates in each of the $k$ partitions, we can expect the CPU utilization to be $\frac{1}{c}$.

% %The total number of edge cuts between partitions are given by $E_c(G,k)$. 
%  So the computational complexity per superstep is given by the complexity for largest subgraphs, $\mathcal{O}_c(\frac{|V|+|E|+\chi(G, k)}{k})$. As remote messages are passed across remote edges, the remote communication complexity per superstep is $\mathcal{O}_m^r(\chi(G,k))$.

% For both FP and HP size of largest subgraph will be close $\frac{|V|}{k \times c}$. As a result computational complexity per superstep reduces to $\mathcal{O}_c(\frac{e+v+\chi^{FP}(G, k\times c)}{k\times c}))$ and $\mathcal{O}_c(\frac{e+v+\chi^{HP}(G, k\times c)}{k\times c}))$ for FP and HP respectively. Similarly the communication cost for FP and HP increase to $\mathcal{O}_m^r(\chi^{FP}(G,k \times c))$ and $\mathcal{O}_m^r(\chi^{HP}(G,k \times c))$. 

% \subsubsection{Makespan}

\subsection{Analysis for Breadth First Search (BFS)}
\label{sec-ana-bfs}
BFS starts at a source vertex and traverses neighboring vertices, one level at a time, marking each newly visited vertex with its distance from the source, until all vertices in the graph are reached. BFS is a non-stationary algorithm~\cite{mizan}, unlike PR, and hence not all vertices/meta-vertices are active in each superstep. The superstep count is not a constant either. Here, we offer an analytical model for the number of supersteps, communication and computational complexity for BFS.
%For Analysis of BFS we are ignoring the communication cost as only $mathcal{O}(E)$ messages are sent across all supersteps.\drnote{only true for vertex centric}

\begin{figure*}[t!]
\centering%~%
  \subfloat[\emph{5 Nodes}]{\includegraphics[width=0.40\textwidth]{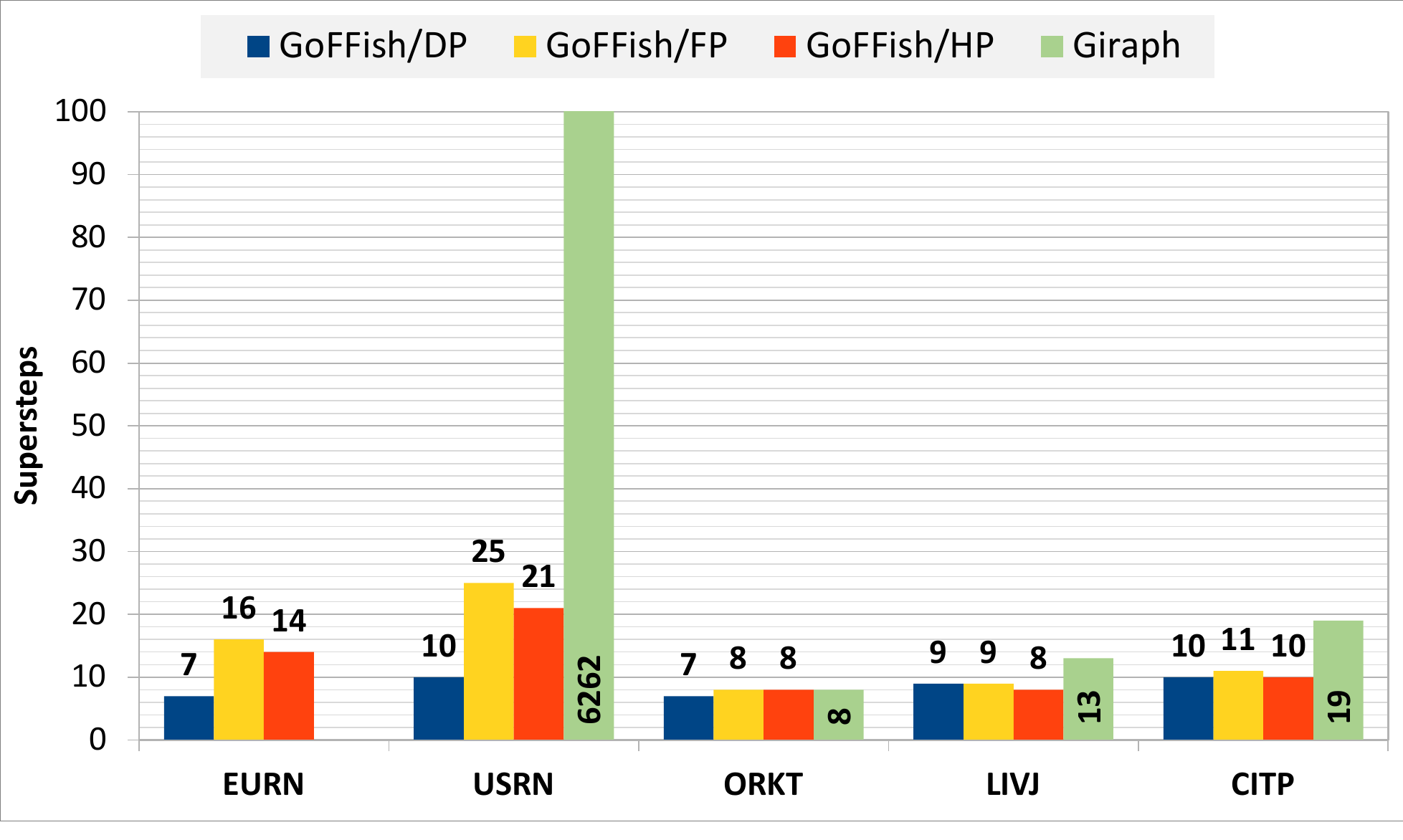}
    \label{fig:sstep:bfs:5m}
  }~~~~
  \subfloat[\emph{10 Nodes}]{\includegraphics[width=0.40\textwidth]{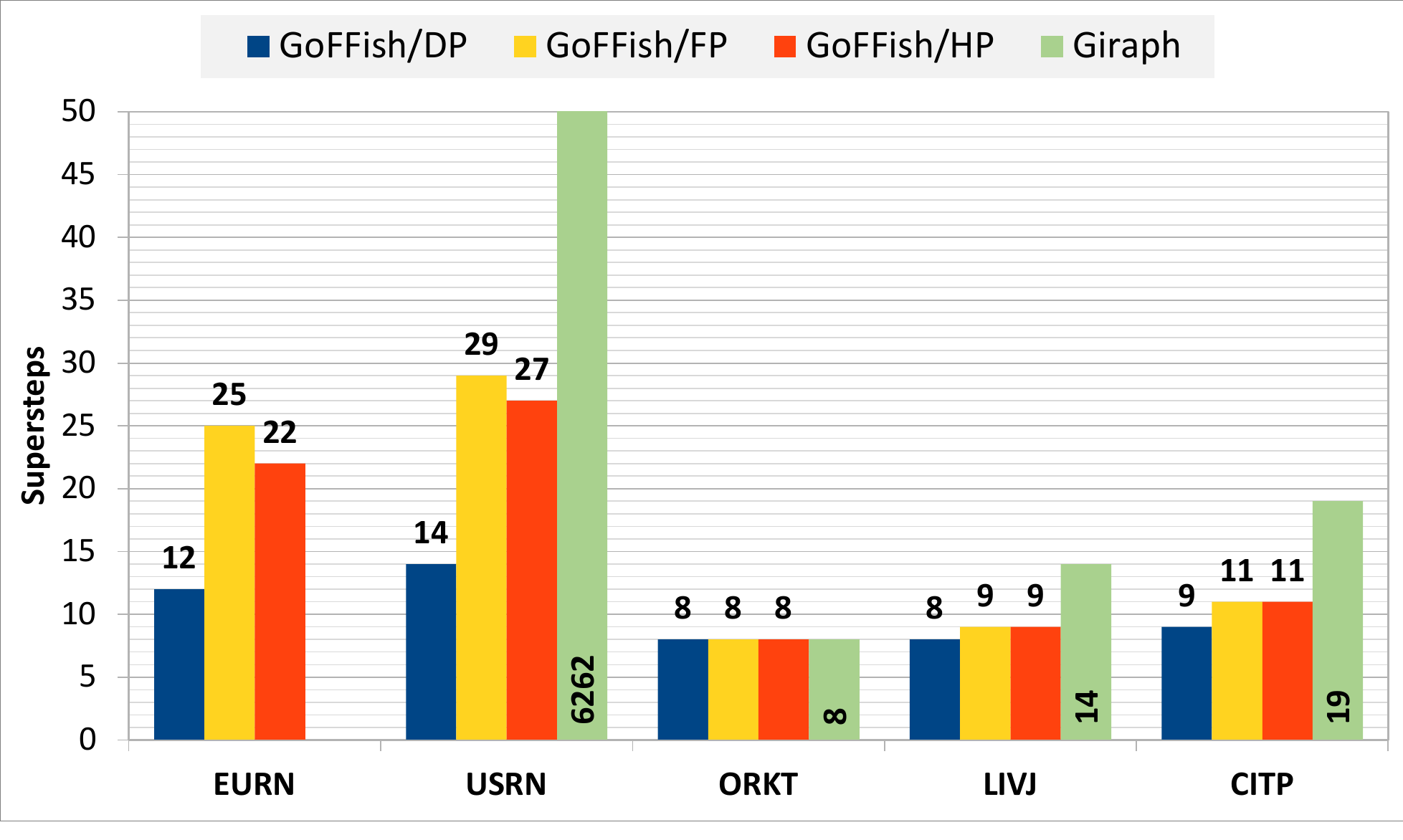}
    \label{fig:sstep:bfs:10m}
  }
\caption{Total supersteps to perform BFS. USRN values are trimmed for display, and EURN not feasible for Giraph in time.}
\label{fig:sstep:bfs}
\vspace{-0.1in}
\end{figure*}

\subsubsection{Number of Supersteps}
In a vertex-centric model, BFS progresses by one level in each superstep as each vertex can only talk to its neighbor. As a result, the \emph{upper bound} on the number of supersteps required to complete the BFS is given by the diameter of the original graph. For e.g., Fig.~\ref{fig:sstep:bfs:5m} shows the number of supersteps taken by Giraph to perform BFS on the five graphs. We note that EURN could not complete BFS on Giraph even after running it for {3~hours}. So its value is not reported. In all other cases, the number of supersteps taken is close to and less than the diameter of the graph, as given in Table~\ref{tbl:graphs}. While for USRN, this value is identical, $6262$, for both 5 and 10 machines, CITP takes $13$ and $14$ supersteps for 5 and 10 machines, given a diameter of $16$. Spatial networks USRN and EURN require a large number of supersteps compared to powerlaw graphs because of their large diameters.

\begin{figure*}[!t]
\centering%~%
  \subfloat[\emph{5 Nodes} (EURN values trimmed for display)]
  {\includegraphics[width=0.40\textwidth]{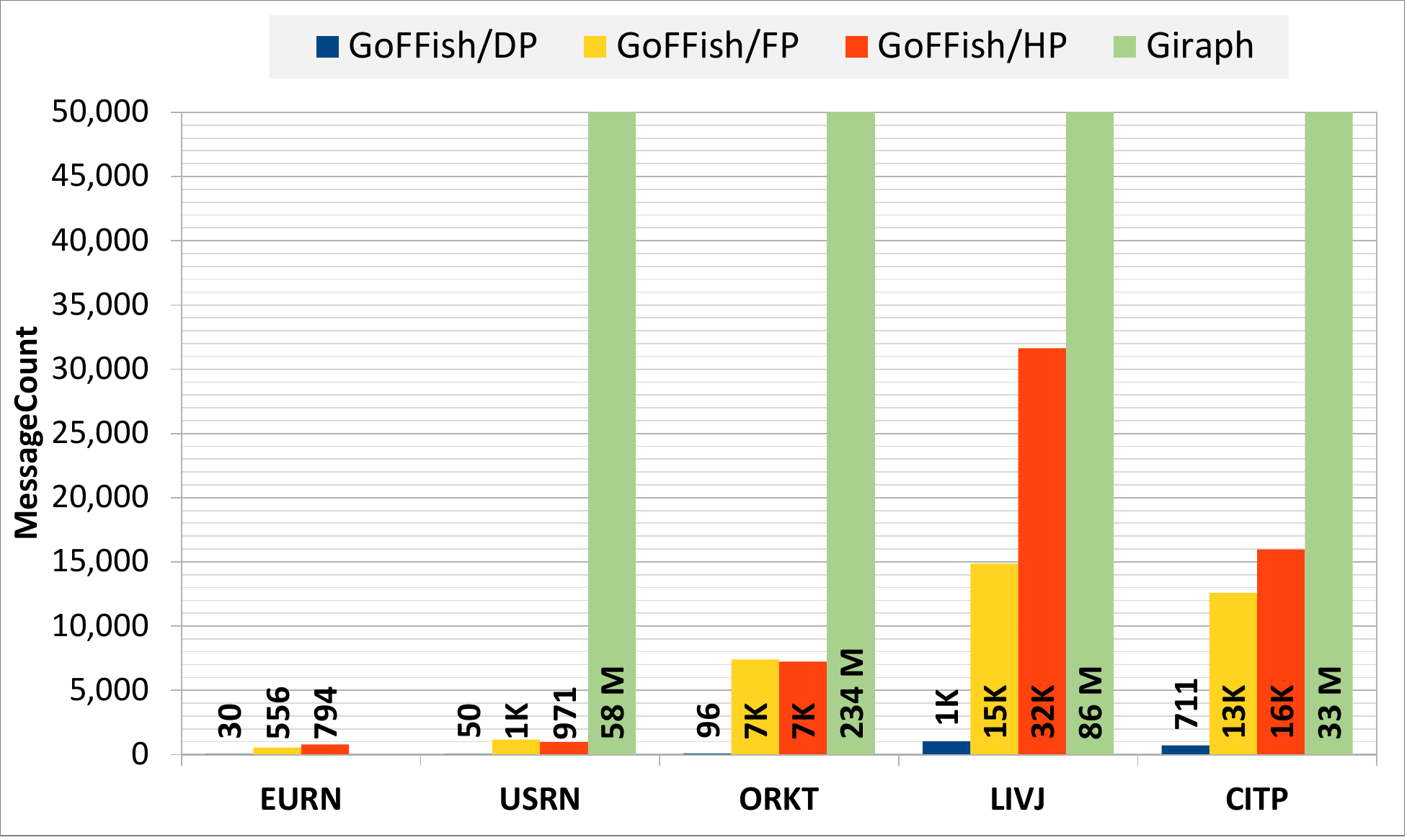}
    \label{fig:msgs:bfs:5m}
  }
  \subfloat[\emph{10 Nodes} (EURN \& LIVJ values trimmed for display)]{\includegraphics[width=0.40\textwidth]{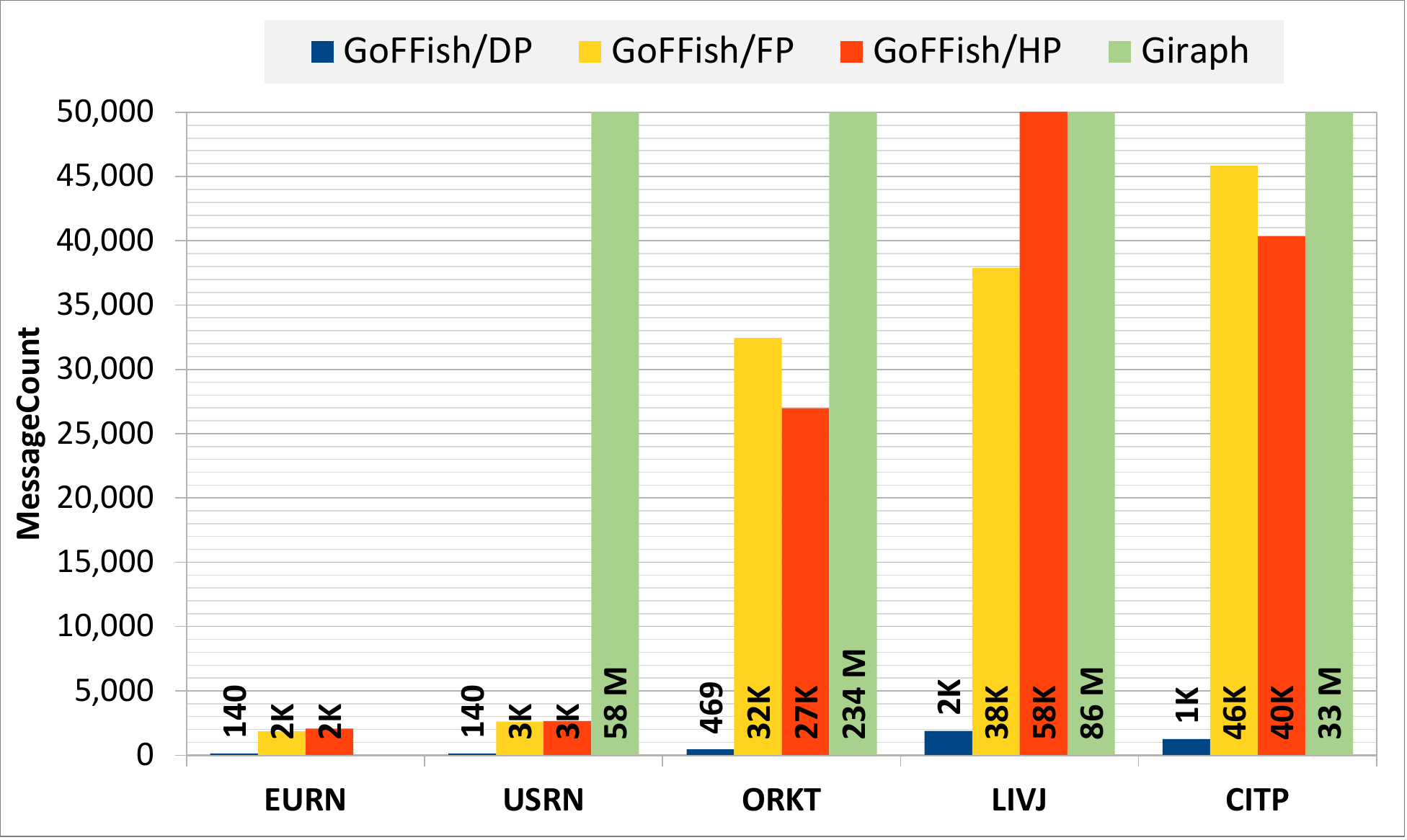}
    \label{fig:msgs:bfs:10m}
  }
\caption{Total number of messages exchanged between components during the BFS.}
\label{fig:msgs:bfs}
\vspace{-0.1in}
\end{figure*}

\begin{figure*}[!ht]
\centering%~%
  \subfloat[\emph{5 Nodes}]{\includegraphics[width=0.40\textwidth]{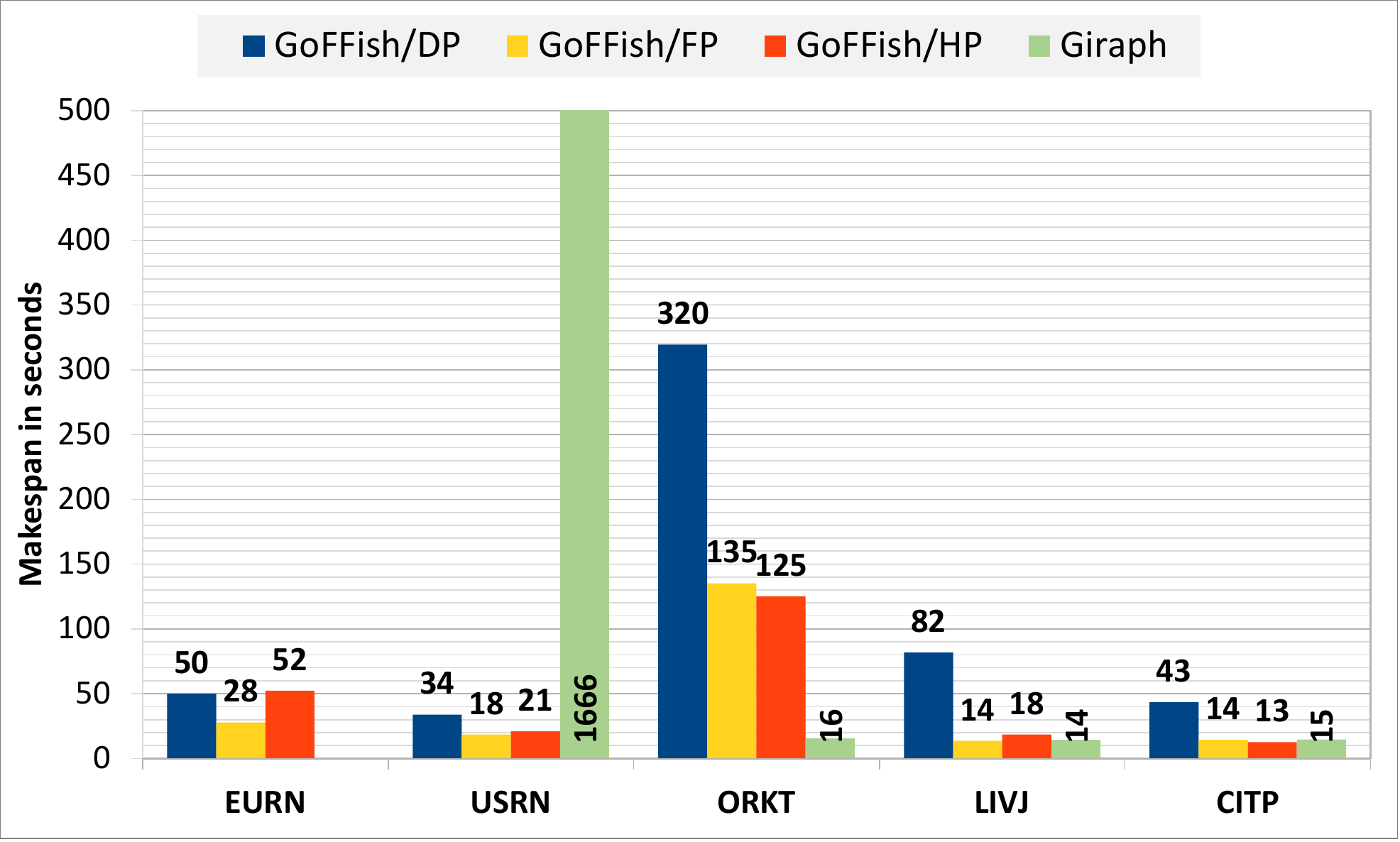}
    \label{fig:Makespan-SSSP-5M}
  }
  \subfloat[\emph{10 Nodes}]{\includegraphics[width=0.40\textwidth]{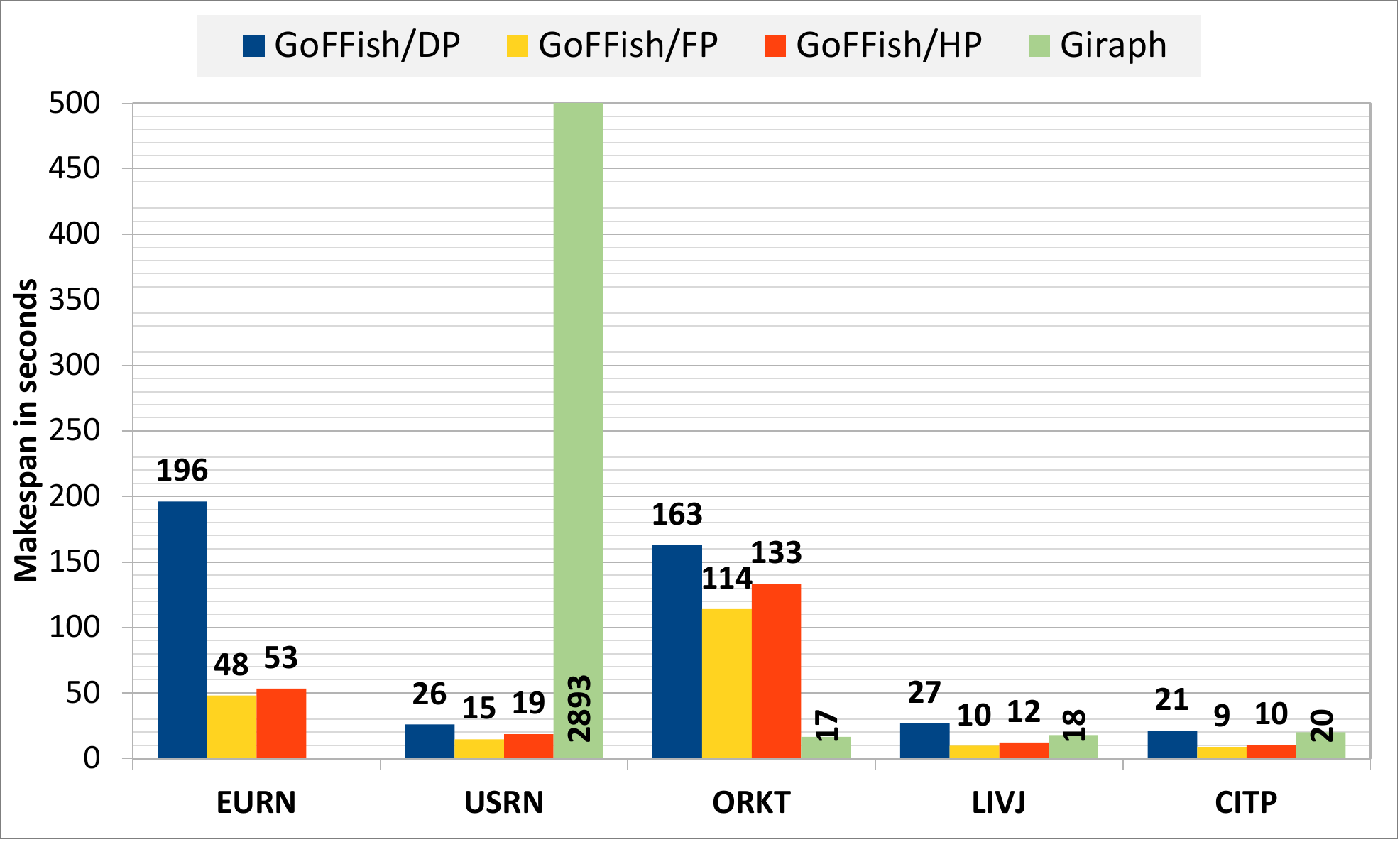}
    \label{fig:Makespan-SSSP-10M}
  }
\caption{Makespan for BFS. USRN values are trimmed for display, and EURN not feasible for Giraph in time.}
\label{fig:makespan-SSSP}
\vspace{-0.1in}
\end{figure*}
For the subgraph-centric models, the BFS algorithm first performs an \emph{A-Star}~\cite{astar,goffish} shortest path algorithm on each subgraph in a superstep, then sends updated shortest distances from vertices that have changed distances to remote subgraphs they are connected to. Vertices receiving new distances form the update vertex set from where \emph{A-Star} is repeated in the next superstep, and so on until all vertex distances stabilize.

In this model, the number of supersteps will be equal to the maximum number of meta-edges that are traversed in any path between the source vertex and every other vertex in the original graph.
%
% Let $sup$ be the total number supersteps taken to complete BFS starting from source $s$ and $L_s$  be the longest shortest path to any $v_i \in V$ from source $s$.  For BFS in subgraph centric model $sup$ is the largest number of times a meta edge $\widehat{e_i} \in \widehat{E}$ is touched in a path $p$ from the source s to any vertex $v_i \in V$ in the BFS tree created from source $s$ in the original graph. 
%
The \emph{absolute lower bound} of the number of supersteps is the radius of the meta-graph, i.e., the minimum eccentricity of any meta-vertex, while the \emph{expected lower bound} is the diameter of the meta-graph. This is apparent, since visiting one vertex in each meta-vertex will require a traversal on a meta-edge, and the fewest possible such meta-edges is given by the radius, while the upper bound on the radius is the diameter. However, BFS using the subgraph centric model can have \emph{revisits}~\cite{dindokar:parlearning:2015}, where the shortest path between two vertices in the same subgraph is through a different subgraph. As a result, a \emph{loose upper bound} for the number of supersteps is the diameter of the whole graph. But this occurs only when every edge in the longest path in the graph occurs on a meta-edge. We see that typically, the number of supersteps is close to the lower bound of the diameter of the meta-graph. 
 
Fig.~\ref{fig:sstep:bfs} shows the number of supersteps taken by BFS for  different graphs and partitioning strategies using GoFFish. Here, we see that the number of supersteps taken is close to the diameter of their respective graphs shown in Table~\ref{tbl:graphs}, and also follow a similar relative trend across partitioning techniques. For e.g., BFS on USRN with 5 machines takes $14$, $29$ and $27$ supersteps using DP, FP and HP, while the diameters of their respective meta-graphs are $4$, $11$ and $11$, or about $3\times$ larger than their meta-graph diameter but much smaller than the USRN graph diameter of $6262$. BFS on CITP with 10 machines takes 10, 11 and 10 supersteps for DP, FP and HP, for meta-graph diameters of $3$ in all these cases, again about $3\times$ more and also fewer than the graph diameter of 22. Since powerlaw graphs have smaller diameters, the upper and lower bounds of the supersteps are closer than for spatial graphs.

 % For spatial graph $\chi(G,k) \lll e$ and $diameter(G)$ (as well as $|L_s|$) is large. As $|\widehat{E}| \leqslant \chi(G,k)$ , the probability of all edges of path $L_s$ in meta edges is extremely small. Same holds for other large shortest paths starting from source $s$. As a result in general cases $sup \lll \mathcal{O}(diameter(G)$ and we see a sharp reduction in number of supersteps from vertex centric model.
 
 % For power law graphs $diameter(G)$ is small and $\chi(G,k)$ is large for higher values of k. Consequently $sup$ can be close to  $diameter(G)$ for larger values of k.

%\begin{figure*}[!ht]
%\centering%~%
%  \subfloat[\emph{5 Nodes} (EURN values trimmed for display)]
%  {\includegraphics[width=0.40\textwidth]{Message-SSSP-5M.PDF}
%    \label{fig:msgs:bfs:5m}
%  }
%  \subfloat[\emph{10 Nodes} (EURN \& LIVJ values trimmed for display)]{\includegraphics[width=0.40\textwidth]{Message-SSSP-10M.PDF}
%    \label{fig:msgs:bfs:10m}
%  }
%\caption{Total number of messages exchanged between components during the BFS.}
%\label{fig:msgs:bfs}
%\vspace{-0.1in}
%\end{figure*}

%\begin{figure*}[!ht]
%\centering%~%
%  \subfloat[\emph{5 Nodes}]{\includegraphics[width=0.40\textwidth]{Makespan-SSSP-5M.pdf}
%    \label{fig:Makespan-SSSP-5M}
%  }
%  \subfloat[\emph{10 Nodes}]{\includegraphics[width=0.40\textwidth]{Makespan-SSSP-10M.pdf}
%    \label{fig:Makespan-SSSP-10M}
%  }
%\caption{Makespan for BFS. USRN values are trimmed for display, and EURN not feasible for Giraph in time.}
%\label{fig:makespan-SSSP}
%\vspace{-0.1in}
%\end{figure*}

\subsubsection{Communications Complexity}
In a vertex-centric model, the total number of messages that will be passed between vertices across all supersteps will equal the number of edges $|E|$, since one traversal message will go on every edge. The number of remote messages, however, will vary based on the number of partitions and will be the number of edge cuts in the partitioned graph.  As observed before, the number of edge cuts and hence, the \emph{expected number of remote messages}, is $\mathcal{O}_m((1 - \frac{1}{p})\cdot |E|)$. Fig.~\ref{fig:msgs:bfs} shows the total number of messages passed by Giraph across all supersteps, and we can see that this matches the number of edges in the graph. EURN could not complete on Giraph in reasonable time, and is not shown.

For the subgraph-centric model, the \emph{lower bound} on the number of messages is the number of meta-edges. This in intuitively similar to the discussion on the number of supersteps: to perform a BFS, every vertex in every meta-vertex has to be visited at least once, and this can be done only by traversing each meta-edge once. However, a \emph{tight upper bound} is harder to pin down since the number of revisits of a subgraph is hard to model. Empirically, based on Fig.~\ref{fig:msgs:bfs}, we see that the communication complexity is $\mathcal{O}_m(\alpha \cdot |\widehat{E}|)$, where the $\alpha$ scaling factor is $\approx 3 - 5\times$. This means that each meta-edge on average has 3--5 physical messages pass through in across supersteps, which means, each meta-vertex is revisited that many times.

We see this correlation when we compare the number of meta-edges in Table~\ref{tbl:graphs}, and the number of physical messages passed for the 5 graphs and 3 partitioning strategies, when running BFS using GoFFish (Fig.~\ref{fig:msgs:bfs}). We see two things: in all cases, the number of messages are between $3\times$ ($50$ for USRN/5M/DP) to $5\times$ ($1159$ for USRN/5M/FP) the number of meta-edges ($16$ and $216$). We also see that the relative trends between the number of meta-edges are seen in the relative number of messages as well. For e.g., ORKT on 10 machines has messages for DP $<$ HP $<$ FP, which is the same trend as their meta-edge count, and likewise, LIVJ on 10 machines has message and meta-edge count that follow DP $<$ FP $<$ HP.

%
%\begin{figure*}[ht]
%\centering%~%
%  \subfloat[\emph{5 Nodes}]{\includegraphics[width=0.40\textwidth]{Makespan-SSSP-5M.pdf}
%    \label{fig:Makespan-SSSP-5M}
%  }
%  \subfloat[\emph{10 Nodes}]{\includegraphics[width=0.40\textwidth]{Makespan-SSSP-10M.pdf}
%    \label{fig:Makespan-SSSP-10M}
%  }
%\caption{Makespan for BFS. USRN values are trimmed for display, and EURN not feasible for Giraph in time.}
%\label{fig:makespan-SSSP}
%\vspace{-0.1in}
%\end{figure*}

\subsubsection{{Computational Complexity}}

Fig.~\ref{fig:makespan-SSSP} the makespan for the different graphs and machines for running BFS using GoFFish and Giraph. The complexity of BFS depends on the size of its frontier set. Let the \emph{frontier} vector $dist_G$ for $G$ have $dist_G[i]$ represent the number of vertices which are of distance $i$ from the source $s$. In a \emph{vertex-centric model}, $dist_G[x]$ vertices are active frontiers at superstep $x$, and all edges of the active vertices will be processed and send messages in next superstep. The frontier set of vertices will initially increase until it reaches a peak size and decreases again~\cite{beamer2013direction}.  With a perfect hash partition, we will get $\mathcal{O}(\frac{dist_G[x]}{p}))$ active vertex on each core for superstep $x$, except for initial and tail supersteps.
 
 For spatial graphs which have uniform degree distribution, the total computational complexity per core for superstep $x$ can be given by $\mathcal{O_c}(\frac{dist_G[x] \times \overline{d}}{p})$ where $\overline{d}$ is the mean degree of vertices in the spatial graph. However, due to skewed degree distribution in power law graphs, the computational complexity will vary across cores. This is because a partition with high degree vertex will process a large number of incoming messages and send large number of outgoing message~\cite{graphlab}.

For DP using a subgraph-centric approach, we get $k$ large meta-vertices each having $\mathcal{O}(\frac{n}{k})$ vertices. This determines the computational complexity per superstep, given by $\mathcal{O}(\frac{n}{k} \times log(\frac{n}{k}))$. Moreover in case of BFS the active vertices moves in wave from the subgraph with the source vertex $s$. As a result the  utilization  for initial supersteps is even worse than $\frac{1}{c}$. This improves as all the meta-vertices become active. % utilization stabilises to $\frac{1}{c}$
For HP and FP the largest subgraph per partition can be of size $\mathcal{O}(\frac{n}{k \times c})$. So the computational complexity reduces to $\mathcal{O}(\frac{n}{k \times c} \times log(\frac{n}{k \times c}))$. Even though for the initial supersteps the utilization remains low due to small number of active subgraphs, utilization increase considerably when all subgraphs become active.

% We know from earlier sections that  both $\chi^{FP}(G,k\times c)  >  \chi(G,k)$ and  $\chi^{HP}(G,k\times c)  >  \chi(G,k)$.
% Consequently the expected number of superstep increases. The growth is number of edge cuts is linear for spatial graph and it decays out for power law graph for large values of k. As a result the rise in expected number of supersteps in case of spatial graphs should be higher. Also we know that in general $\chi^{HP}(G,k\times c) > \chi^{FP}(G,k\times c)$ for near optimum partition strategy. So the number of supersteps in case of HP should be higher than FP.

% \section{Discussion}
% \ysnote{Summary of results and observations, from Sec 4 and 5, in bullets}

\section{Conclusion}%~~~\note{0.75 page}}
\label{sec:conclusion}
% \Note{Conclusion unfinished. Doesn't address footnote 2.}\\
% \Note{sufficient system architectural considerations in designing the strategies [out of scope, Java ForkJoinPool]}

In this paper, we have formalized the concept of meta-graphs as an analytical sketch over large graphs. We see the distinctive nature of the $3$ partitioning algorithms for spatial and powerlaw graphs, and, further small-world graphs within them. We offer a sound analytical basis to examine their impact on the meta-graph's number of meta-vertices, meta-edges, weights and diameter, and further validate these using the meta-graph statistics. Meta-graphs also exhibit a recursive behavior of their structure with the original graph, indicating their suitability as a coarse-grained approximation for analysis.

We use this meta-graph sketch to analyze and offer bounds on the communication and computational complexity for PR and BFS graph algorithms using a subgraph-centric model, and complement this with bounds for the vertex-centric model as well. While in some cases the bounds are tight, such as the supersteps for BFS, in others, like the communication cost for BFS, these are relaxed. These are also validated against experimental results from both GoFFish and Giraph platforms.

These methods offer a formal foundation to examine the effectiveness of component-centric programming models to support Big Graph applications.  As part of future work, the meta-graph sketch can also be used for other novel analysis, such as understanding the behavior of non-stationary algorithms for elastic scheduling~\cite{ccgrid}. There is also scope to tighten these bounds and offer formal proofs where possible. 

\bibliographystyle{IEEEtran}
\footnotesize{
\bibliography{paper}
}

\end{document}